\documentclass[twocolumn]{aa520}
\usepackage{psfig}
\usepackage{epsfig}
\usepackage{longtable}
\usepackage{lscape}
 
\def\kms{$\mathrm{km\;s}^{-1}$} 
\def\dg{^\circ} 
\def\ha{H$\alpha$}
\def\hb{H$\beta$}

\def\oiiipg{[O~{\small III}]$\,\lambda\lambda4959,5007$}

\def\niipg{[N~{\small II}]$\,\lambda\lambda6548,6583$}

\def\siipg{[S~{\small II}]$\,\lambda\lambda6716,6731$} 

\def\kmsmpc{$\rm km\;s^{-1}\;Mpc^{-1}$}

\def\h3{$h_{3}$}
\def\h4{$h_{4}$}

\begin{document} 
 
\title{Minor-axis velocity gradients in disk galaxies\thanks{Based on
    observations carried out at the European Southern Observatory in
    La Silla (Chile) (ESO 69.B-0706 and 70.B-0338), with the Multiple
    Mirror Telescope which is a joint facility of the Smithsonian
    Institution and the University of Arizona, and with the Italian
    Telescopio Nazionale Galileo (AOT-5, 3-18) at the Observatorio del
    Roque de los Muchachos in La Palma (Spain).}$^{\bf,}$\thanks{Table 5 is only available in electronic format the CDS via anonymous
    ftp to cdsarc.u-strasbg.fr (130.79.128.5) or via
    http://cdsweb.u-strasbg.fr/Abstract.html. Table 1 is only available in electronic form at {\tt http://www.edpsciences.org}}}

\author{L. Coccato\inst{1}
 \and E.~M. Corsini\inst{1}
 \and A. Pizzella\inst{1}
 \and L. Morelli\inst{1,2}
 \and J.~G. Funes S.~J.\inst{3}
 \and F. Bertola\inst{1}}
 
\offprints{coccato@pd.astro.it}
  
\institute{
Dipartimento di Astronomia, Universit\`a di Padova, 
  vicolo dell'Osservatorio~2, I-35122 Padova, Italy \and
European Southern Observatory, 3107 Alonso de Cordova, 
  Santiago, Chile \and 
Vatican Observatory, University of Arizona, Tucson, AZ 85721, USA} 

\date{Version: \today }  
  
\titlerunning{Minor-axis velocity gradients in disk
    galaxies}
\authorrunning{Coccato et al.}

\abstract{We present the ionized-gas kinematics and photometry of a sample of 4
  spiral galaxies which are characterized by a zero-velocity plateau
  along the major axis and a velocity gradient along the minor axis,
  respectively. By combining these new kinematical data with those
  available in the literature for the ionized-gas component of the S0s
  and spirals listed in the Revised Shapley-Ames Catalog of Bright
  Galaxies we realized that about $50\%$ of unbarred galaxies show a
  remarkable gas velocity gradient along the optical minor axis. This
  fraction rises to about $60\%$ if we include unbarred galaxies with
  an irregular velocity profile along the minor axis. This phenomenon is
  observed all along the Hubble sequence of disk galaxies, and it is
  particularly frequent in early-type spirals. Since minor-axis
  velocity gradients are unexpected if the gas is moving onto circular
  orbits in a disk coplanar to the stellar one, we conclude that
  non-circular and off-plane gas motions are not rare in the inner
  regions of disk galaxies.
}

\maketitle

\section{Introduction}

The ionized-gas velocity curves measured along the disk major axis are
commonly adopted to derive the mass distribution of spirals within
their optical region (see Sofue \& Rubin 2001 for a review).
In the past decades the central velocity gradient of the gas rotation
curves has been used to constrain the amount and distribution of the
visible component in high surface brightness galaxies (e.g. Kent 1986;
Persic, Salucci \& Stel 1996). During recent years a lively debate
concerning the dark matter distribution in the central regions of low
surface brightness galaxies has been triggered by the measurement of
the inner slope of the ionized-gas rotation curves in a large sample
of these galaxies (McGaugh, Rubin
\& de Block 2001; de Block, McGaugh \& Rubin 2001).

These mass models are based on the hypothesis that ionized gas in
central regions is moving onto circular orbits and in the plane of the
stellar disk. However, evidence has mounted that the kinematic
behavior of the gaseous component in inner regions of disk galaxies is
more complex.
Indeed ionized gas can have an intrinsic velocity dispersion and its
velocity can be lower than the circular speed derived from dynamical
models based on the stellar kinematics and photometry (Fillmore,
Boroson \& Dressler 1986; Bertola et al. 1995; Cinzano et al. 1999), a
large number of kinematically-decoupled gaseous components rotating in
a different plane with respect to that of the stellar disk have been
found in early-type disk galaxies (Bertola, Buson \& Zeilinger 1992;
Bertola \& Corsini 2000; Corsini et al. 2003), and non-circular gas
motions have been observed in triaxial bulges (Bertola, Rubin \&
Zeilinger 1989; Gerhard et al. 1989; Berman 2001). Moreover the vast
majority of the gas velocity fields measured by Hubble Space Telescope
in the nuclei of nearby spiral galaxies are not regular and possibly
affected by non-gravitational forces (Sarzi 2003).
This poses new questions about the reliability of gas kinematics to
derive the mass density profile in the central regions of galaxies.

The presence of a velocity gradient along the disk minor axis is the
kinematic signature that ionized gas is not moving onto circular
orbits in a disk which is coplanar to that of the stars.
To point out that this is a common phenomenon we have built a
compilation of the minor-axis velocity profiles available from
long-slit spectroscopy for the disk galaxies listed in the Revised
Shapley-Ames Catalog of Bright Galaxies (Sandage \& Tammann 1981, RSA
hereafter).  In addition, we have obtained the minor-axis velocity
profiles for 4 spiral galaxies, whose major-axis kinematics are
characterized by a remarkable zero-velocity plateau. All these
galaxies show a gas velocity gradient along the disk minor axis. Such
a peculiar kinematics may be also the signature of an inner polar disk
as those we have recently found in early-type spiral galaxies (Corsini
et al. 2003).

This paper is organized as follows. The list of the RSA disk
galaxies for which the ionized-gas velocity profile has been
measured along the disk minor axis is given in
Sec. \ref{sec:catalog}. The spectroscopic observations and data
analysis of our 4 sample galaxies are described in
Sect. \ref{sec:observations}. The resulting ionized-gas kinematics are
discussed in Sect. \ref{sec:kinematics}. Our conclusions are presented
in Sect. \ref{sec:conclusions}.

\section{Minor-axis velocity profiles of ionized gas of 
  the disk galaxies in the Revised Shapley-Ames Catalog}
\label{sec:catalog}
 
To get an exhaustive picture of the phenomena related to
non-circular and off-plane motions of ionized gas in disk galaxies, we
compiled in Table 1 both data from the literature and
from the present paper. In Table 1 we listed all the
disk galaxies of RSA for which the ionized-gas velocity profiles
has been measured along the minor axis by means of long-slit
spectroscopy. The table give the main properties of the galaxies,
namely the morphological classification, size, inclination, major-axis
position angle, apparent magnitude and recession velocity. The
position angle of the observed major and minor axis as well as the
related bibliographic reference are reported in Table
1 too.

Minor-axis velocity data have been found in literature for 138 out of
1101 disk galaxies ($12\%$), whose RSA morphological type
ranges from S0 to Sm, including barred systems. During this process we
realized that the velocity curves of the ionized-gas component have a
variety of shapes. For purpose of classification we assigned the
minor-axis velocity profiles to five different classes, which we named
Z, O, C, G, and I, as follows:

\begin{itemize}

\item[{Z:}] A zero velocity gradient is observed for the ionized-gas
  component along the minor axis.

\item[{O:}] An overall gas velocity profile is observed along the minor
  axis without zero-velocity points out to the last measured radius.

\item[{C:}] Non-zero gas velocities are measured only in the central
  regions along the minor axis. The gas velocity profile shows a steep
  gradient rising to a maximum observed velocity of few tens of \kms\
  in the inner few arcsec, then the velocity drops to zero at larger
  radii.

\item[{G:}] A central velocity gradient is observed, but the limited
  radial extension of the data does not allow us to distinguish between
  overall or centrally-confined non-zero velocities.

\item[{I:\ }] There is evidence of non-zero velocities along the minor
  axis but the velocity is either too poorly detected or too
  asymmetric to be assigned to the previous classes.

\end{itemize}

\begin{figure}[ht!]
\centering
\resizebox{\hsize}{!}{ 
\includegraphics[clip=true,angle=0]{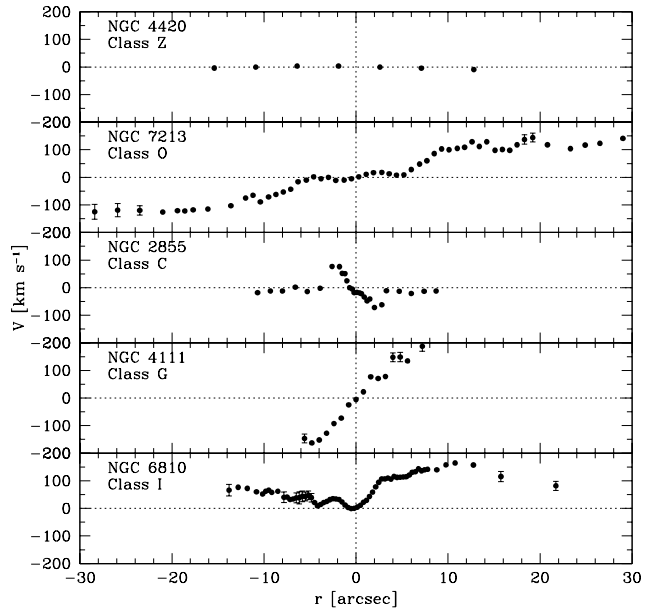}} 
\caption{The ionized-gas velocity profiles measured along the optical
  minor axis of NGC 4420 ($\rm P.A. = 98^\circ$, Fisher et al. 1997),
  NGC 7213 ($\rm P.A. = 34^\circ$, Corsini et al. 2003), NGC 2855
  ($\rm P.A. = 30^\circ$, Corsini et al. 2002), NGC 4111 ($\rm P.A. =
  60^\circ$, Fisher 1997), and NGC 6810 ($\rm P.A. = 86^\circ$, this
  paper) are given as prototypical examples for the five classes of
  minor-axis velocity profiles described in
  Sect. \ref{sec:catalog}. Errorbars smaller than symbols are not
  plotted.}
\label{fig:prototype}
\end{figure}

In Fig. \ref{fig:prototype} we show a prototypic example for each
class of minor-axis velocity profile. The classification of the
minor-axis velocity profiles of RSA galaxies is given in Table
1.

\section{Sample, observations and data reduction}
\label{sec:observations}

NGC 949, NGC 2460, NGC 2541, and NGC 6810 are unbarred spiral galaxies
which belong to a sample of 16 bright ($B_T\leq13.5$) and nearby
($V_\odot\leq3500$ \kms) spirals studied by Coccato et al. (2004) to
search for a circumnuclear Keplerian disk (Bertola et al. 1998; Funes
et al. 2002). The main properties of the sample galaxies are
summarized in Table \ref{tab:sample}.
The major-axis rotation curve of their gaseous component is
characterized by a remarkable zero-velocity plateau. We decided to
obtain spectra along their minor axis to investigate the
possible presence of a velocity gradient. The position angles of
the major and minor axis were chosen according to de Vaucouleurs et 
al. (1991, RC3 hereafter) and therefore relate to the orientation of 
the outermost isophotes. This is confirmed by the isophotal analysis 
of the images available for the sample galaxies in the NASA/IPAC 
Infrared Science Archive of the Two Micron All Sky Survey (2MASS) and 
in the archive by Frei et al. (1996).

\begin{table*}[ht!]
\addtocounter{table}{+1}
\caption{Parameters of the sample galaxies}
\begin{center}
\begin{tabular}{lllccccccc}
\hline
\noalign{\smallskip}
\multicolumn{1}{c}{Object} &
\multicolumn{2}{c}{Morp. Type} &
\multicolumn{1}{c}{$B_T$} &
\multicolumn{1}{c}{P.A.} &
\multicolumn{1}{c}{$i$} &
\multicolumn{1}{c}{$V_{\odot}$} &
\multicolumn{1}{c}{$D$} &
\multicolumn{1}{c}{Scale} &
\multicolumn{1}{c}{$M_{B_T}^0$} \\
\noalign{\smallskip}
\multicolumn{1}{c}{[name]} &
\multicolumn{1}{c}{[RSA]} &
\multicolumn{1}{c}{[RC3]} &
\multicolumn{1}{c}{[mag]} &
\multicolumn{1}{c}{[$\dg$]} &
\multicolumn{1}{c}{[$\dg$]} &
\multicolumn{1}{c}{[\kms]} &
\multicolumn{1}{c}{[Mpc]} &
\multicolumn{1}{c}{[pc/$''$]} &
\multicolumn{1}{c}{[mag]} \\
\noalign{\smallskip}
\multicolumn{1}{c}{(1)} &
\multicolumn{1}{c}{(2)} &
\multicolumn{1}{c}{(3)} &
\multicolumn{1}{c}{(4)} &
\multicolumn{1}{c}{(5)} &
\multicolumn{1}{c}{(6)} &
\multicolumn{1}{c}{(7)} &
\multicolumn{1}{c}{(8)} &
\multicolumn{1}{c}{(9)} &
\multicolumn{1}{c}{(10)} \\
\noalign{\smallskip}
\hline
\noalign{\smallskip}  
\object{NGC~0949}  & Sc(s)III & Sb(rs):?  & 12.40 & 145 & 58 &  608 $\pm10$ & 11.2 &  54.5 & $-18.47$  \\
\object{NGC~2460}  & Sab(s)   & Sa(s)     & 12.72 &  40 & 42 & 1452 $\pm10$ & 21.3 & 103.0 & $-19.35$  \\
\object{NGC~2541}  & Sc(s)III & Scd(s)    & 12.26 & 165 & 60 &  540 $\pm10$ &  8.3 &  40.5 & $-18.03$  \\
\object{NGC~6810}  & Sb       & Sab(s):sp & 12.37 & 176 & 78 & 1961 $\pm5 $ & 24.3 & 117.8 & $-20.44$  \\
\noalign{\smallskip}
\hline
\noalign{\medskip}
\end{tabular}
\end{center}
\begin{minipage}{18cm}
NOTES -- Col.(2): morphological classification from RSA.
Col.(3): morphological classification from RC3.
Col.(4): apparent total blue magnitude from RC3.
Col.(5): major-axis position angle from RC3.
Col.(6): inclination derived as $\cos^{2}{i}\,=\,(q^2-q_0^2)/(1-q_0^2)$.
         The observed axial ratio $q$ is taken from RC3 and the
         intrinsic flattening has been assumed following
         Guthrie (1992) with RC3 morphological classification.
Col.(7): heliocentric velocity of the galaxy derived as center of
         symmetry of the rotation curve of the gas with eliocentric correction.
Col.(8): distance obtained as $V_0/H_0$ with $H_0=75$ \kmsmpc\
         and $V_0$ the systemic velocity derived
         from $V_\odot$ corrected for the motion of the Sun with
         respect of the Local Group as done in the RSA.
Col.(10): absolute total blue magnitude corrected for
          inclination and extinction from RC3.
\end{minipage}
\label{tab:sample}
\end{table*}

\subsection{Kinematic data}

The long-slit spectroscopic observations of the sample galaxies were
carried out at the Roque de los Muchachos Observatory in La Palma with
the Telescopio Nazionale Galileo (TNG), at the European Southern
Observatory in La Silla with the New Technology Telescope (NTT), and
at the Multiple Mirror Telescope Observatory in Arizona with the
Multiple Mirror Telescope (MMT) in 2002.
The details about the instrumental setup of each observing run as well
as the value of the seeing FWHM as measured by fitting a
two-dimensional Gaussian to the guide star are given in
Table~\ref{tab:setup}.

\begin{table*}[!ht]
\caption{Setup of the spectroscopic observations.}
\begin{small} 
\begin{tabular}{lcccc} 
\hline
\noalign{\smallskip}
Parameter  & Run 1 &  Run 2 & Run 3 & Run 4\\
\noalign{\smallskip}
\hline 
\noalign{\smallskip}  
Date         & Jan 19--21, 2002 & May 16, 2002 & Oct 08, 2002 &  Oct 31--Nov 01, 2002 \\
Instrument   & TNG$+$DOLORES & NTT$+$EMMI & NTT$+$EMMI & MMT$+$Blue Channel\\ 
Grating/Grism& VPH \#9 1435 $\mathrm{gr\;mm}^{-1}$ & \#6 1200 $\mathrm{gr\;mm}^{-1}$ & 
  \#7 600 $\mathrm{gr\;mm}^{-1}$ & 1200 $\mathrm{gr\;mm}^{-1}$ \\ 
CCD          & $2048\times2048$ Loral & \#36 $2048\times2048$ Tek & $2048\times4096$ MIT/LL & \#22 $3072\times1024$ Loral \\
Pixel size   & $15\times15$ $\mathrm{\mu\,m}^2$ & $24\times24$ $\mathrm{\mu\,m}^2$ & 
  $15\times15$ $\mathrm{\mu\,m}^2$ & $15\times15$ $\mathrm{\mu\,m}^2$ \\
Scale        & $\mathrm{0\farcs275\;pixel}^{-1}$ & $\mathrm{0\farcs32\;pixel}^{-1}$ & 
  $\mathrm{0\farcs332\;pixel}^{-1}$ & $\mathrm{0\farcs30\;pixel}^{-1}$ \\ 
Dispersion   & $\mathrm{0.34\;\AA\;pixel}^{-1}$ & $\mathrm{0.32\;\AA\;pixel}^{-1}$ & 
  $\mathrm{0.83\;\AA\;pixel}^{-1}$ & $\mathrm{0.49\;\AA\;pixel}^{-1}$ \\ 
Slit width   & $1\farcs0$  &  $1\farcs0$ & $1\farcs0$  &  $1\farcs5$ \\ 
Wavelength range  & 6210--6870 \AA\ &  6400--7020 \AA\ & 5630--6970 \AA\  &   4455--5945 \AA\ \\
Instrumental FWHM  &  1.63 \AA\ & 1.12 \AA\ & 2.34 \AA\ &  2.02 \AA\ \\ 
Instrumental $\sigma^{\mathrm a}$  &  $32$ \kms\ & $22$ \kms\ & $45$ \kms\ &  $53$ \kms \\ 
Seeing FWHM  &  $1\farcs5$--$2\farcs0$ & $1\farcs8$ & $1\farcs3$ & $1\farcs8$--$2\farcs2$ \\ 
\noalign{\smallskip}  
\hline 
\noalign{\smallskip}  
\end{tabular}\\  
\begin{minipage}{8.8cm} 
$^{\mathrm a}$ At \ha\ in Run 1--3, and at \hb\ in Run 4.
\end{minipage} 
\end{small}
\label{tab:setup}
\end{table*}

Medium-resolution spectra were taken along both the major and minor
axis after centering the galaxy nucleus on the slit using the guiding
camera. A comparison spectrum was taken before and/or after every
object exposure. The typical integration time of the galaxy spectra
were 1800 s and 2700 s. Total integration times and slit position
angles of the galaxy spectra as well as the log of observations are
given in Table~\ref{tab:log}.

Basic data reduction was performed as in Corsini et al. (1999).  Using
standard {\tt ESO-MIDAS}\footnote{{\tt MIDAS} is developed and
maintained by the European Southern Observatory.}  routines, all the
spectra were bias subtracted, flat-field corrected by quartz lamp and
twilight exposures, cleaned from cosmic rays, and wavelength
calibrated.
The flat-field correction was performed by means of both quartz lamp
and twilight sky spectra in order to correct for pixel-to-pixel
sensitivity variations and large-scale illumination patterns due to
slit vignetting.
Cosmic rays were identified by comparing the photon counts in each
pixel with the local mean and standard deviation and they were
eliminated by interpolating over. Residual cosmic rays were eliminated
by manually editing the spectra.
We checked that the wavelength rebinning was done properly by
measuring the difference between the measured and predicted
wavelengths (Osterbrock et al. 1996) for the brightest night-sky
emission lines in the observed spectral ranges. The resulting accuracy
in the wavelength calibration is typically of 5 \kms.
Instrumental resolution was derived as the mean of Gaussian FWHM's
measured for a number of unblended arc-lamp lines of a wavelength
calibrated spectrum. The mean FWHM of the arc-lamp lines and the
corresponding resolution at \ha\ (Run 1--3) and \hb\ (Run 4) are given
in Table~\ref{tab:setup}.
After the calibration all the spectra were corrected for CCD
misalignment.  The contribution of the sky was determined from the
outermost $\sim10''$ at the two edges of the resulting frames where
the galaxy light was negligible, and then subtracted.  The spectra
obtained for a given galaxy along the same position angle were coadded
using the center of the stellar-continuum radial profile as reference.

The ionized-gas kinematics was measured by the simultaneous Gaussian
fit of the emission lines present in the spectra (namely \niipg, \ha,
and \siipg\ in Run 1--3, \hb\ and \oiiipg\ in Run 4).  The galaxy
continuum was removed row-by-row by fitting a fourth to sixth order
polynomial avoiding all the spectral regions with strong emission and
absorption features. We fitted in each row of the continuum-subtracted
spectrum a Gaussian to each emission line, assuming them to have the
same line-of-sight velocity and velocity dispersion (corrected for
heliocentric velocity and instrumental FWHM, respectively). In the
spectra an additional absorption Gaussian has been added in the fit to
take into account for the presence of the \ha\ or \hb\ absorption line
and the flux ratio of the \niipg\ lines have been fixed to 1:3. Far
from the galaxy center (for $|r|\ga10''$) we averaged adjacent
spectral rows in order to increase the signal-to-noise ratio of the
relevant emission lines.
Errors on the gas velocity and velocity dispersion were derived from
photon statistics and CCD readout noise by means of Monte Carlo
simulations.

\begin{figure*}[ht!]
\centering
\resizebox{\hsize}{!}{ 
\includegraphics[clip=true,angle=-90]{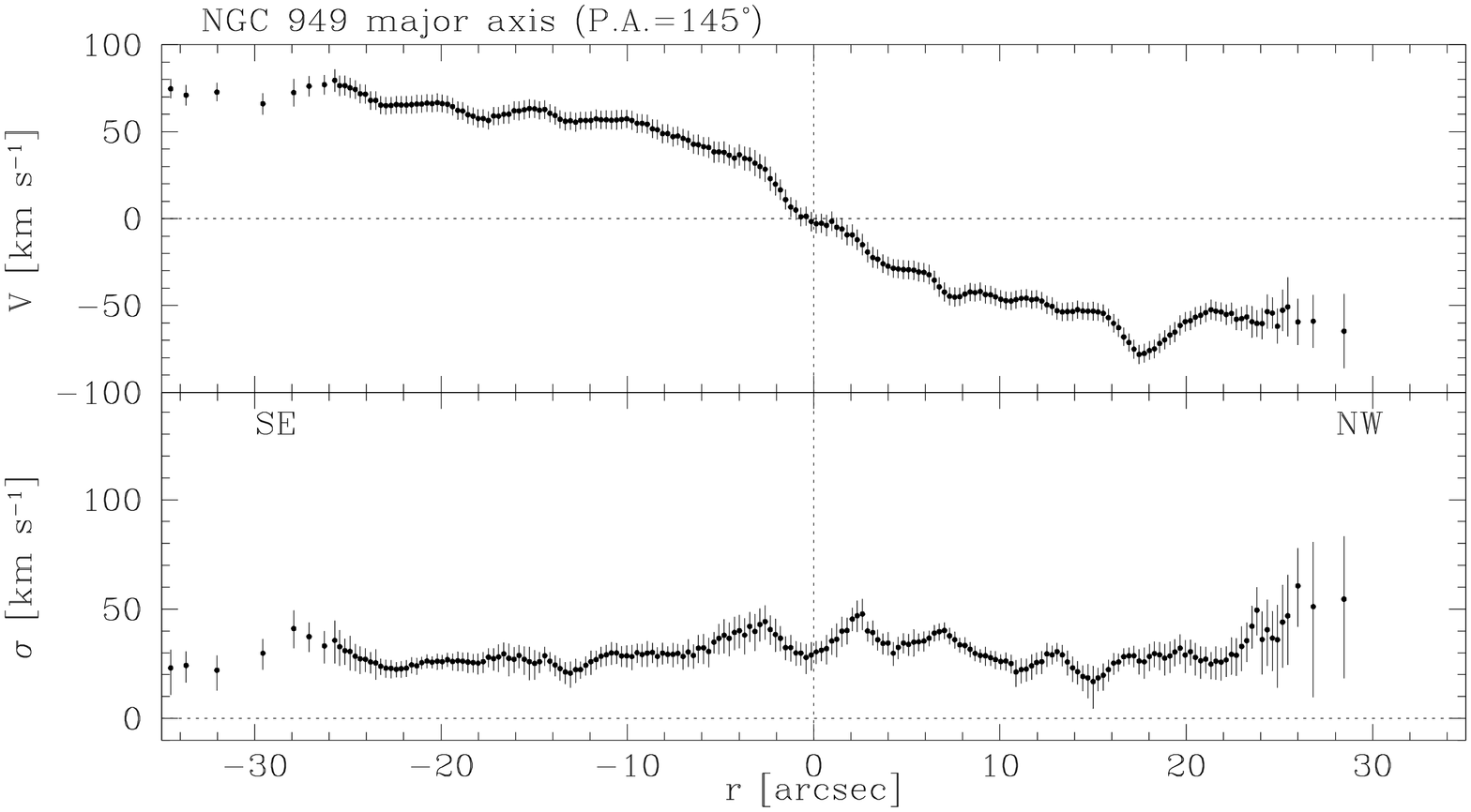} 
\includegraphics[clip=true,angle=-90]{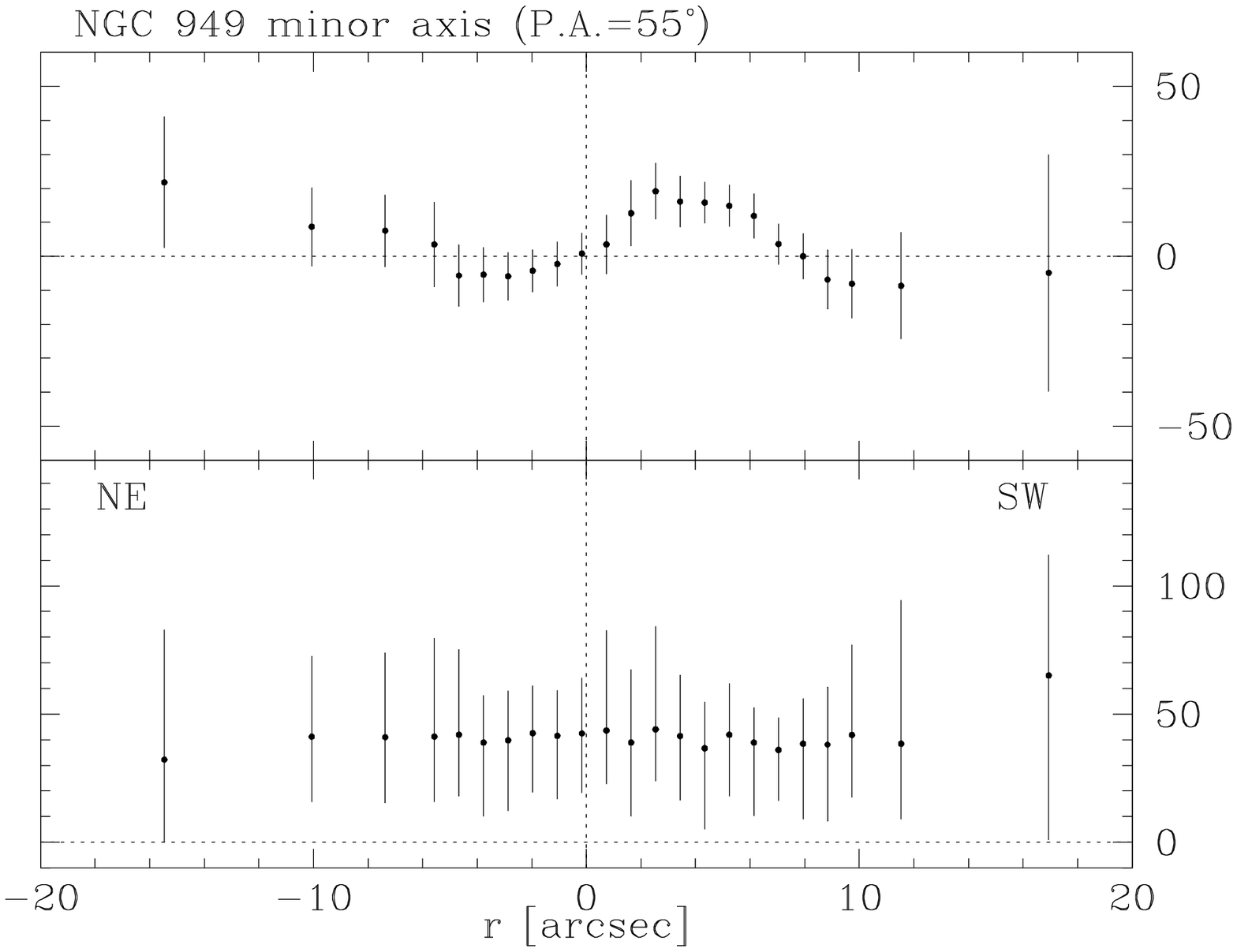}} 
\resizebox{\hsize}{!}{ 
\includegraphics[clip=true,angle=-90]{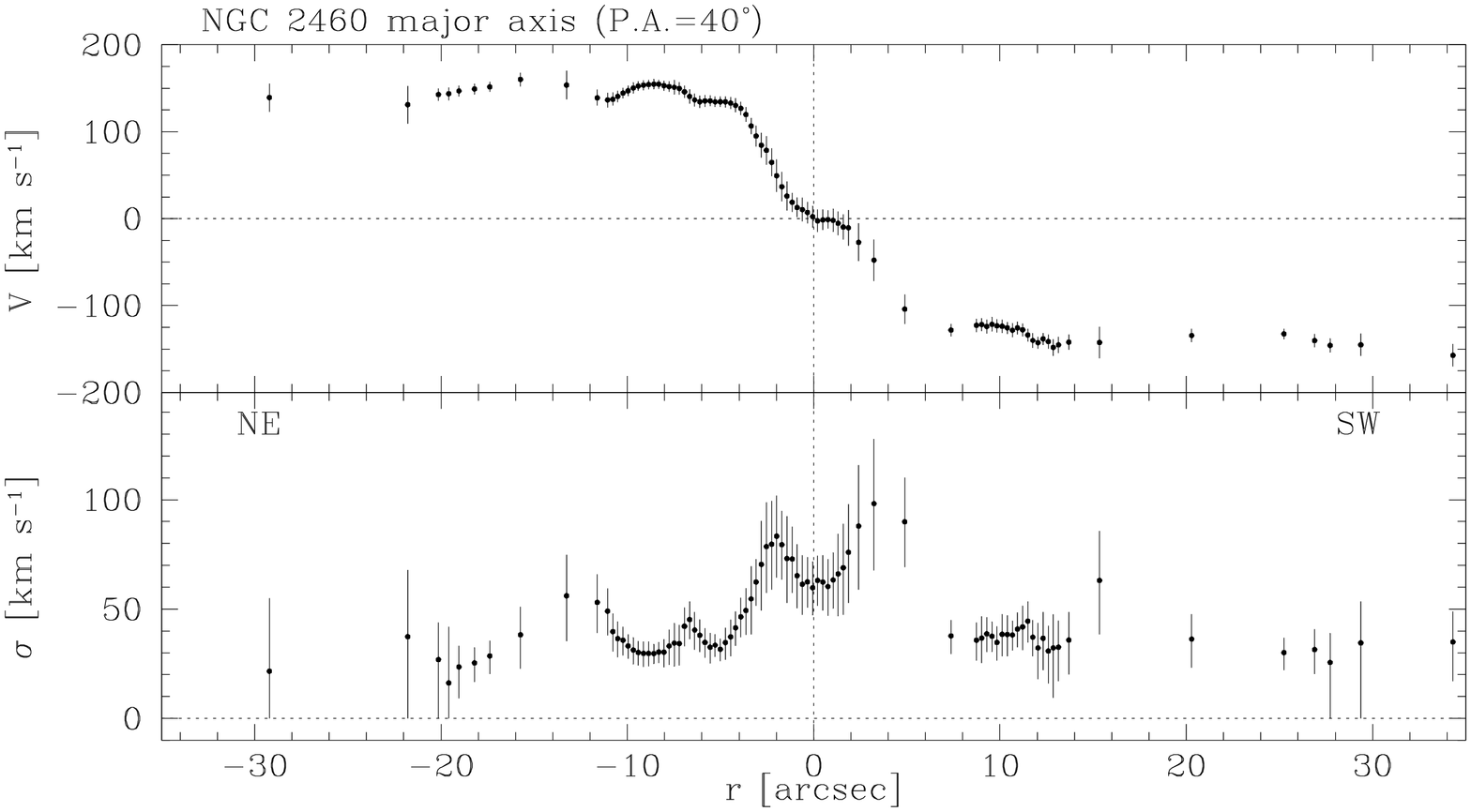} 
\includegraphics[clip=true,angle=-90]{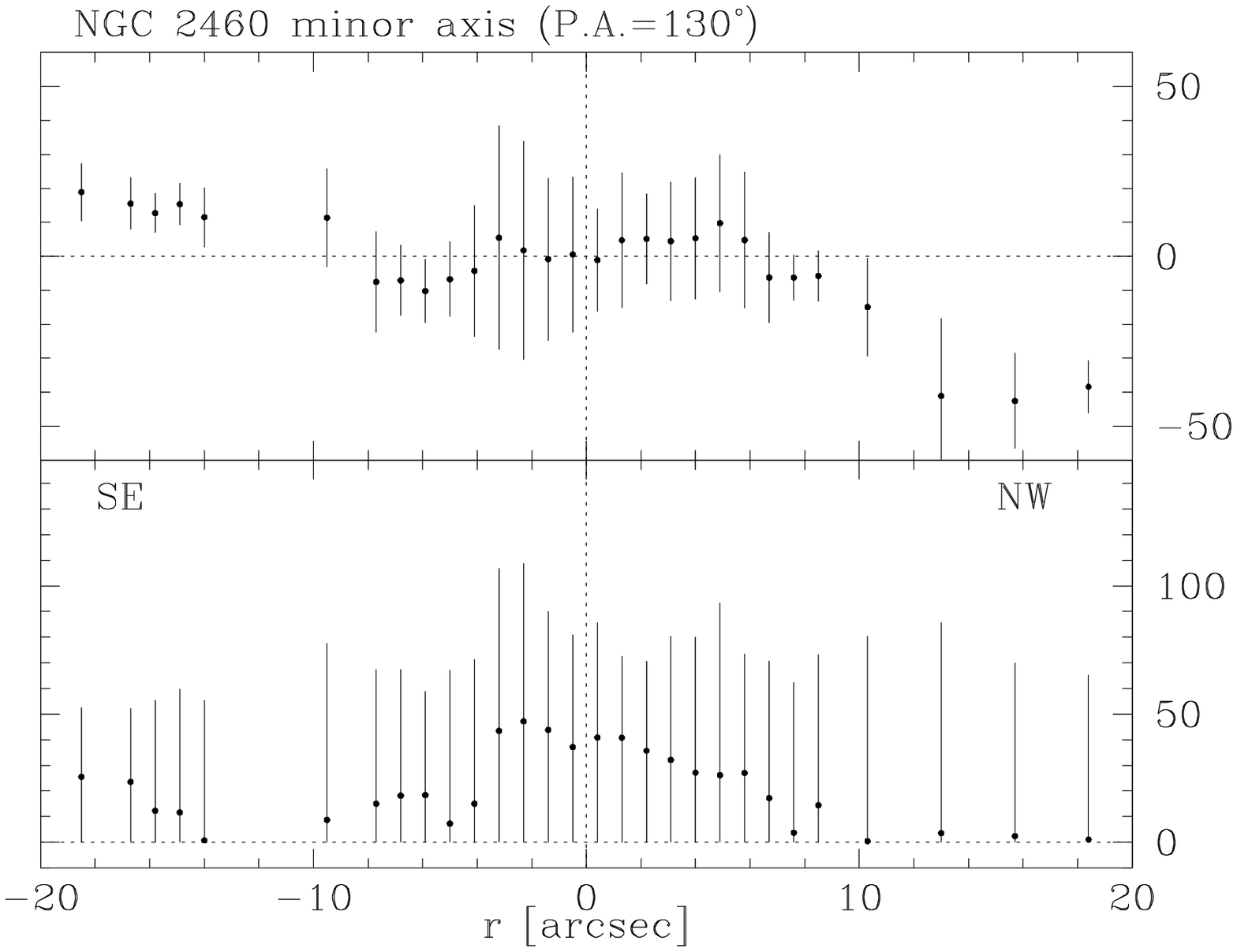}} 
\resizebox{\hsize}{!}{ 
\includegraphics[clip=true,angle=-90]{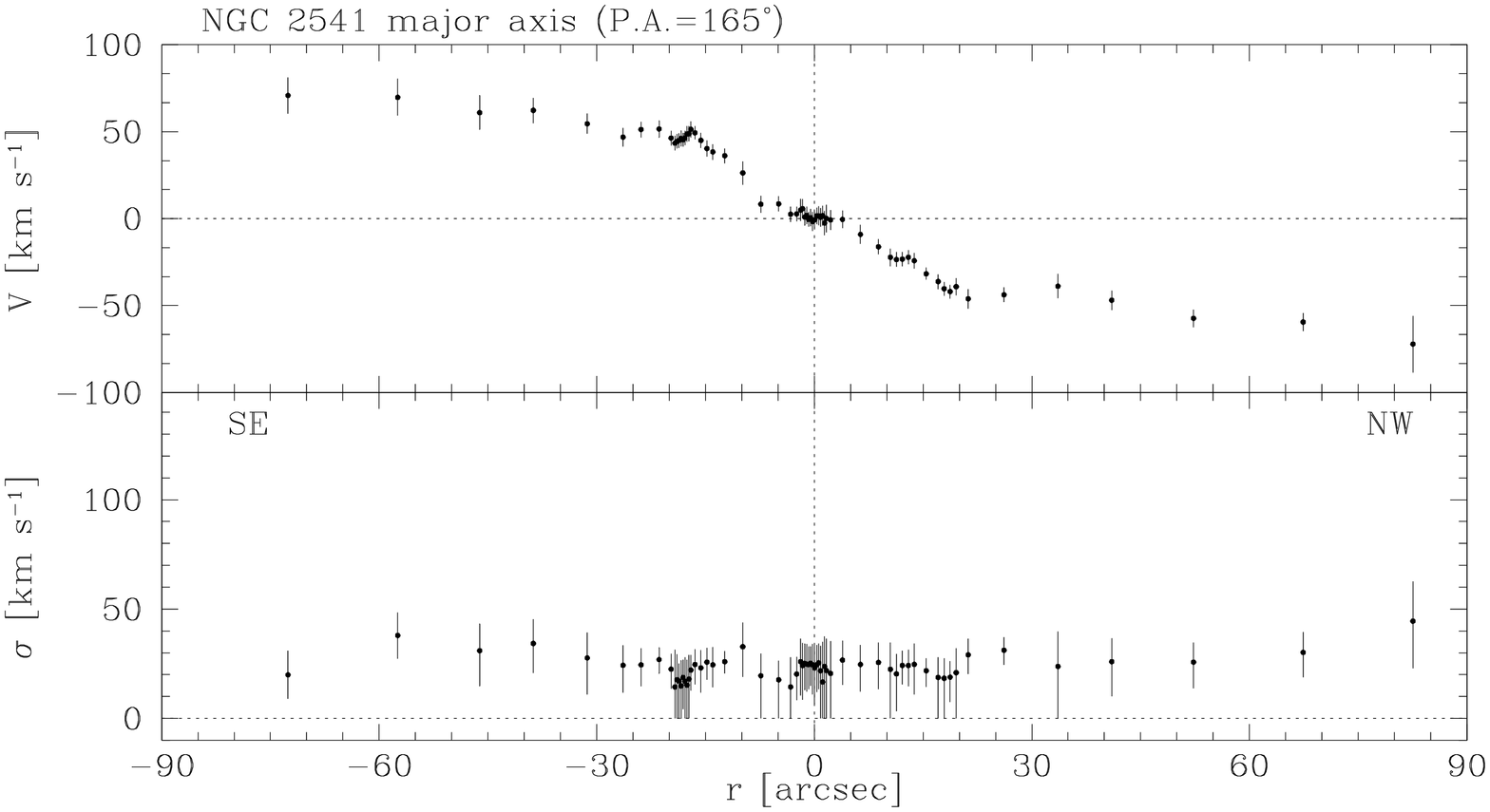} 
\includegraphics[clip=true,angle=-90]{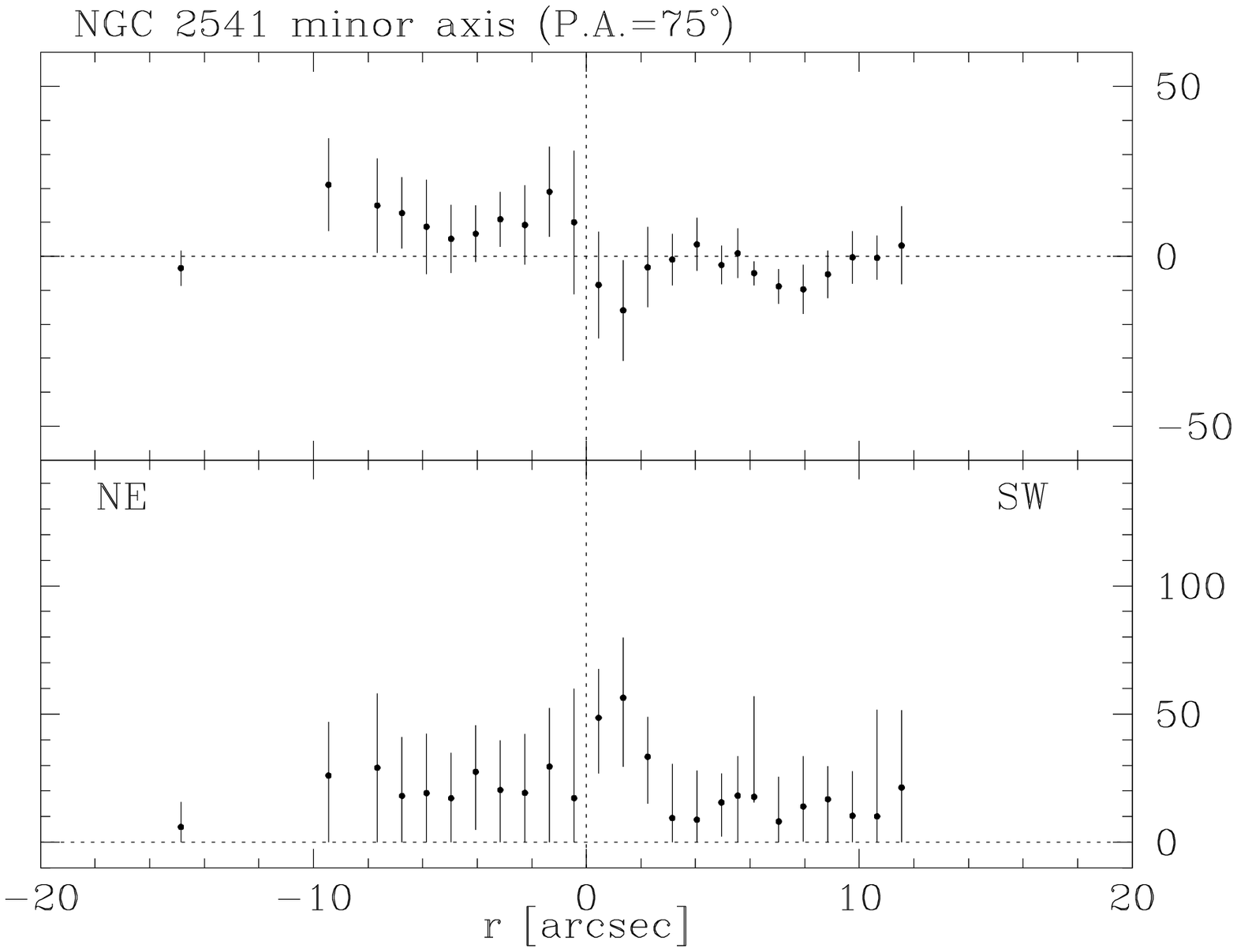}} 
\resizebox{\hsize}{!}{ 
\includegraphics[clip=true,angle=-90]{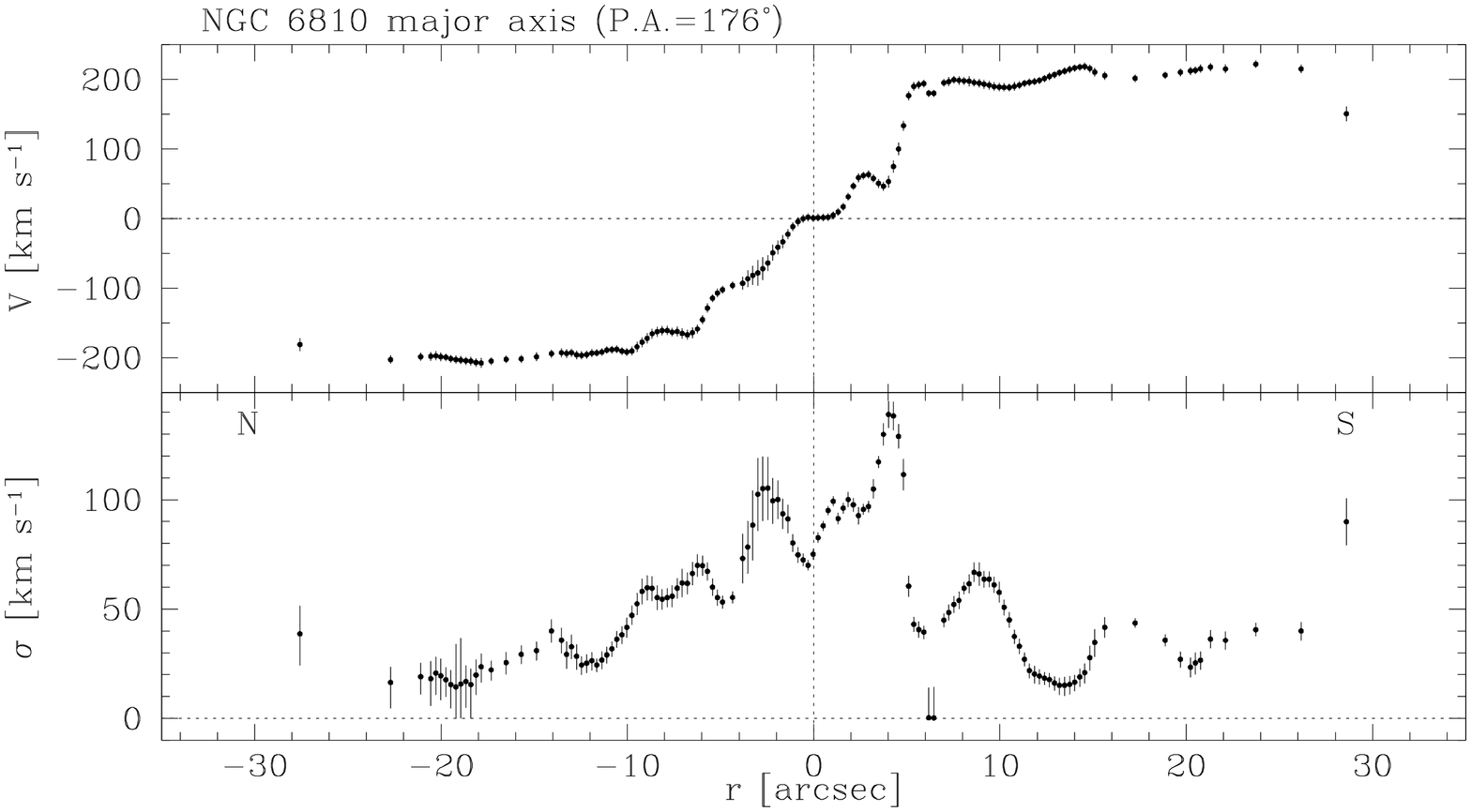} 
\includegraphics[clip=true,angle=-90]{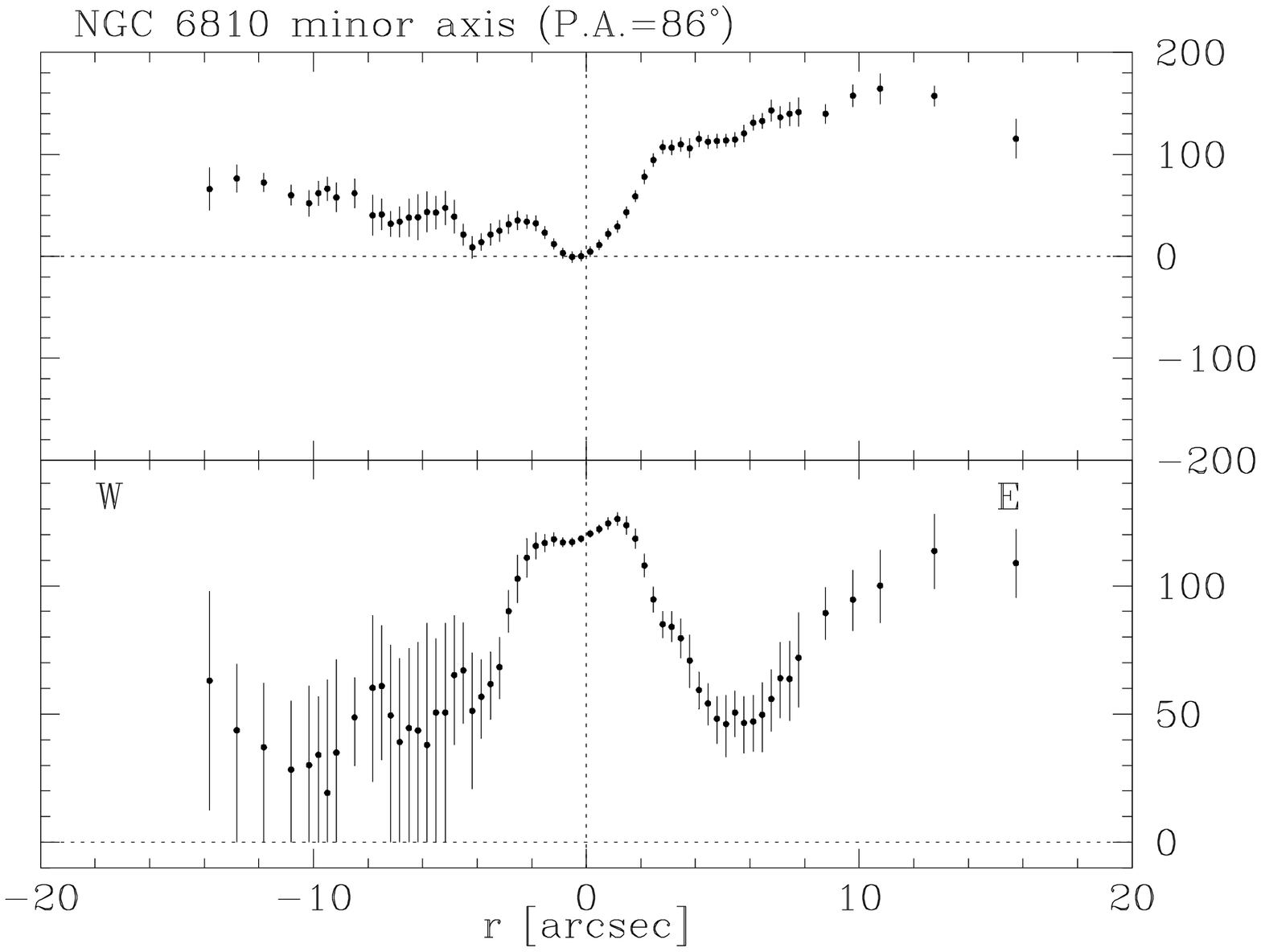}} 
\caption{Ionized-gas kinematics measured along the optical
  major ({\it left panels\/}) and minor axes ({\it right panels\/}) of
  the 4 sample galaxies. Errorbars smaller than symbols are not
  plotted.} 
\label{fig:kinematics}
\end{figure*}

The line-of-sight velocity and velocity dispersion profiles we
measured for the gaseous component along the major and minor axis of
the sample galaxies are presented in Fig. \ref{fig:kinematics} and
values are reported in Table 5. The line-of-sight velocities of
the ionized-gas are the observed ones after subtracting the systemic
velocities of Table \ref{tab:sample} and without applying any
correction for galaxy inclination.

\subsection{Photometric data}

We retrieved the 2MASS $H-$band images of NGC 949
($3\farcm5\times3\farcm5$), NGC 2460 ($3\farcm0\times3\farcm0$), NGC 
2541 ($13\farcm3\times13\farcm3$) and NGC 6810 
($5\farcm0\times5\farcm0$) from the NASA/IPAC Infrared Science 
Archive. The galaxy images were reduced and flux calibrated with the 
standard 2MASS extended source processor GALWORKS (Jarrett et 
al. 2000). Images have a spatial resolution of $1''$ and were obtained 
with a typical seeing ${\rm FWHM\sim1''}$. 

The signal-to-noise of the 2MASS image of NGC 2541 was too low to
perform a reliable photometric analysis, we therefore retrieved the
$i-$band image ($4\farcm0\times4\farcm0$) obtained by Frei et
al. (1996) too. They already performed all the basic steps of data
reduction, including bias subtraction, flat-fielding, cleaning of
cosmic rays, and subtraction of sky background and stellar sources.

\begin{figure*}[ht!]
\begin{center}
\hbox{
\includegraphics[clip=true,width=8cm]{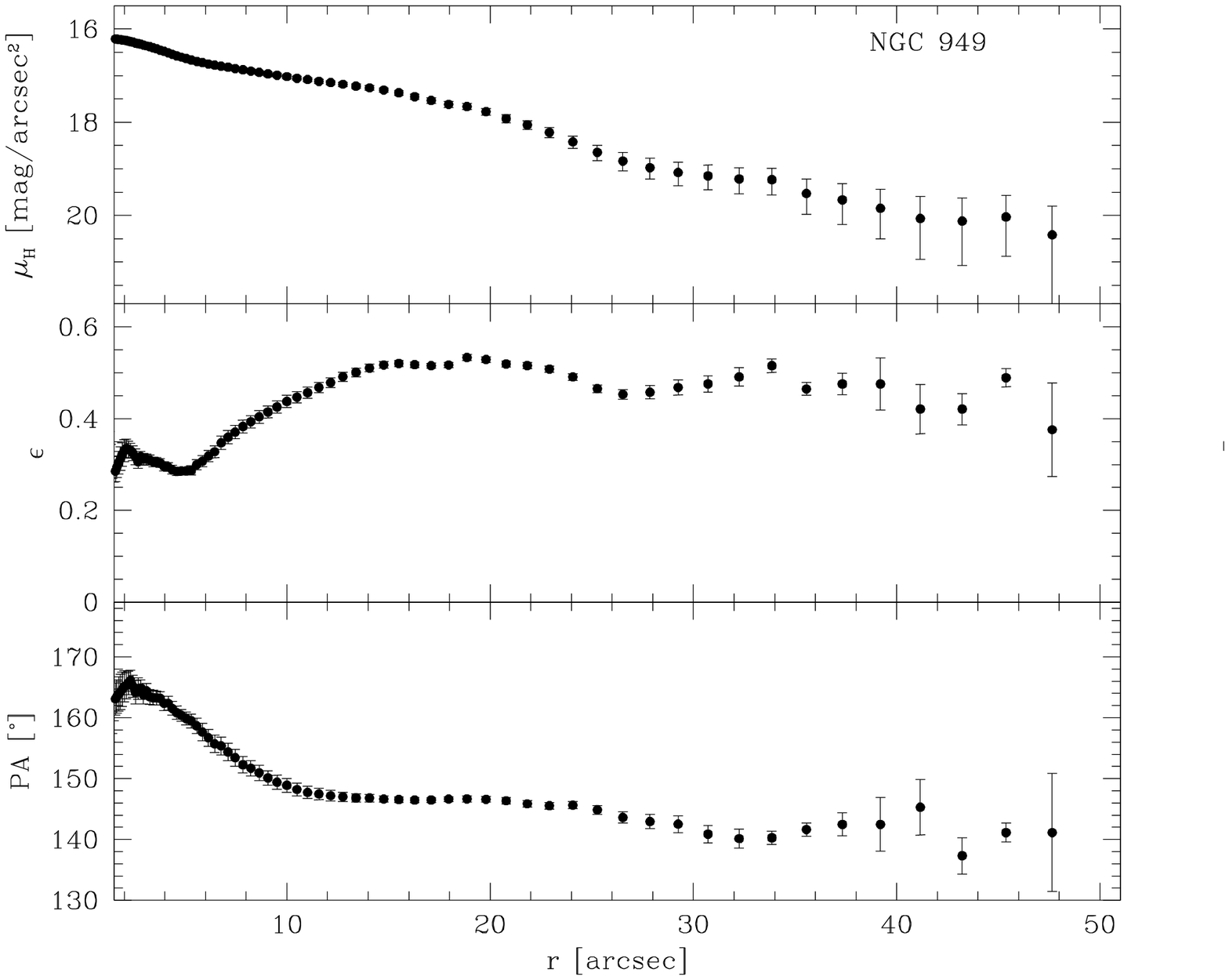} 
\includegraphics[clip=true,width=8cm]{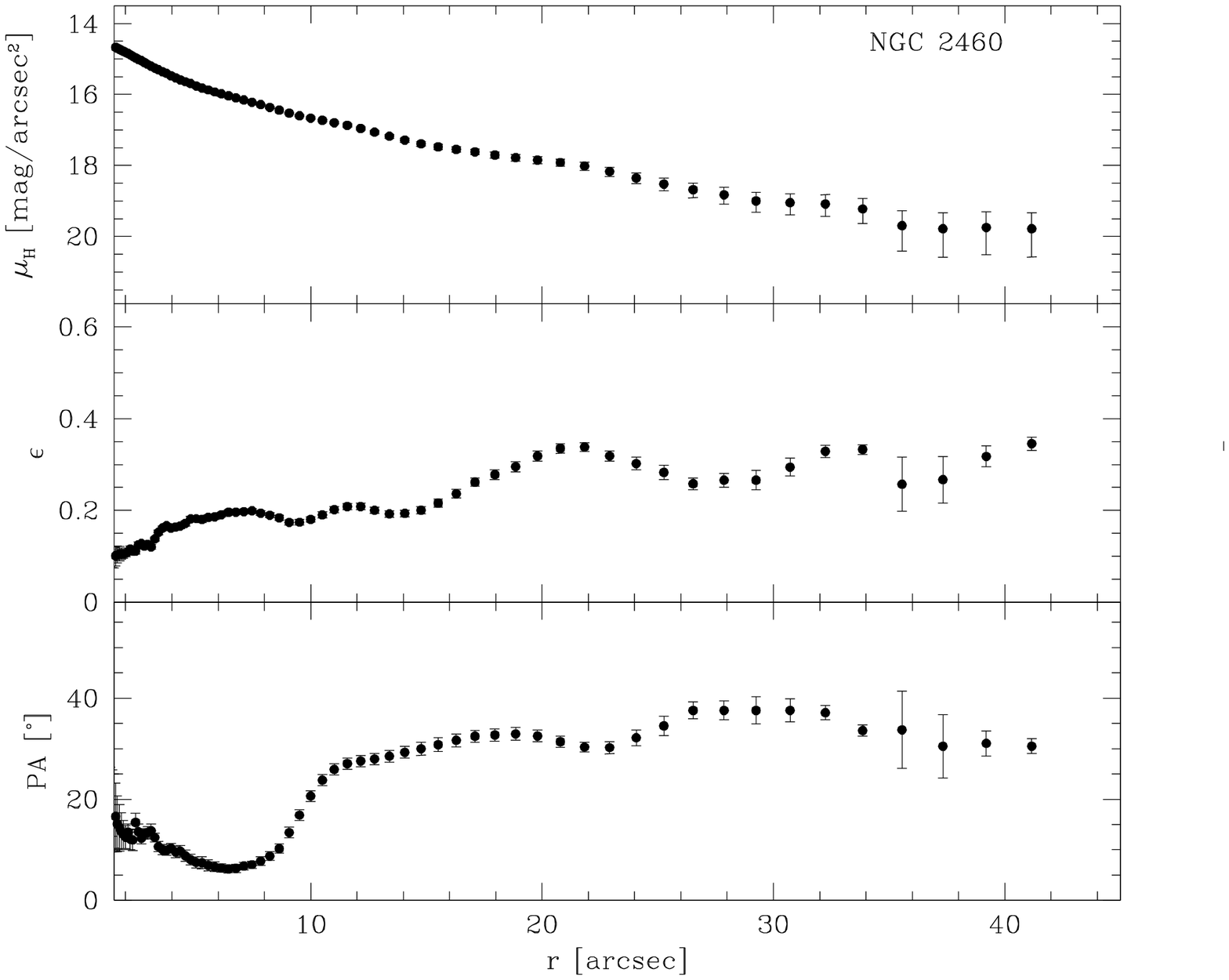}}
\hbox{
\includegraphics[clip=true,width=8cm]{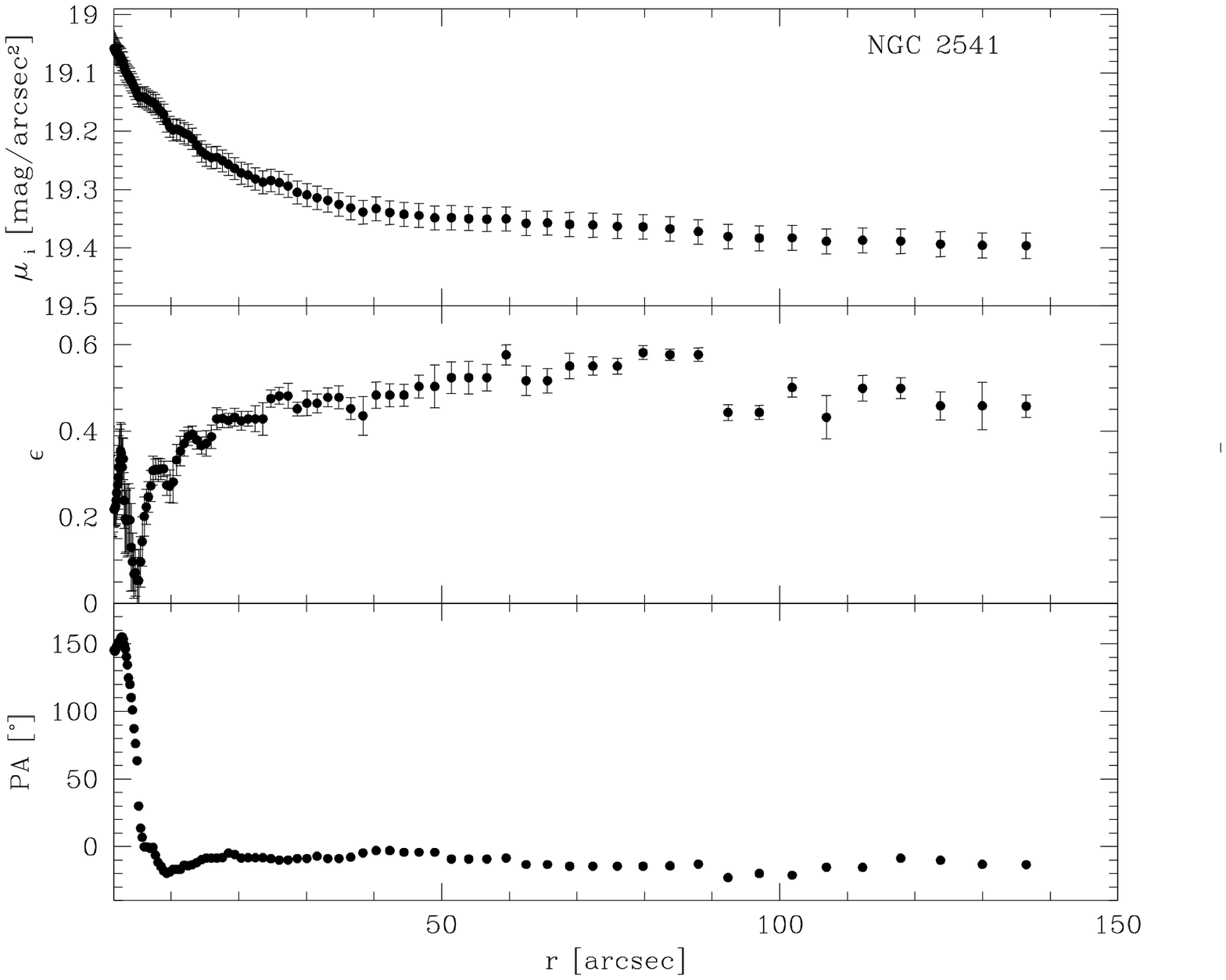}
\includegraphics[clip=true,width=8cm]{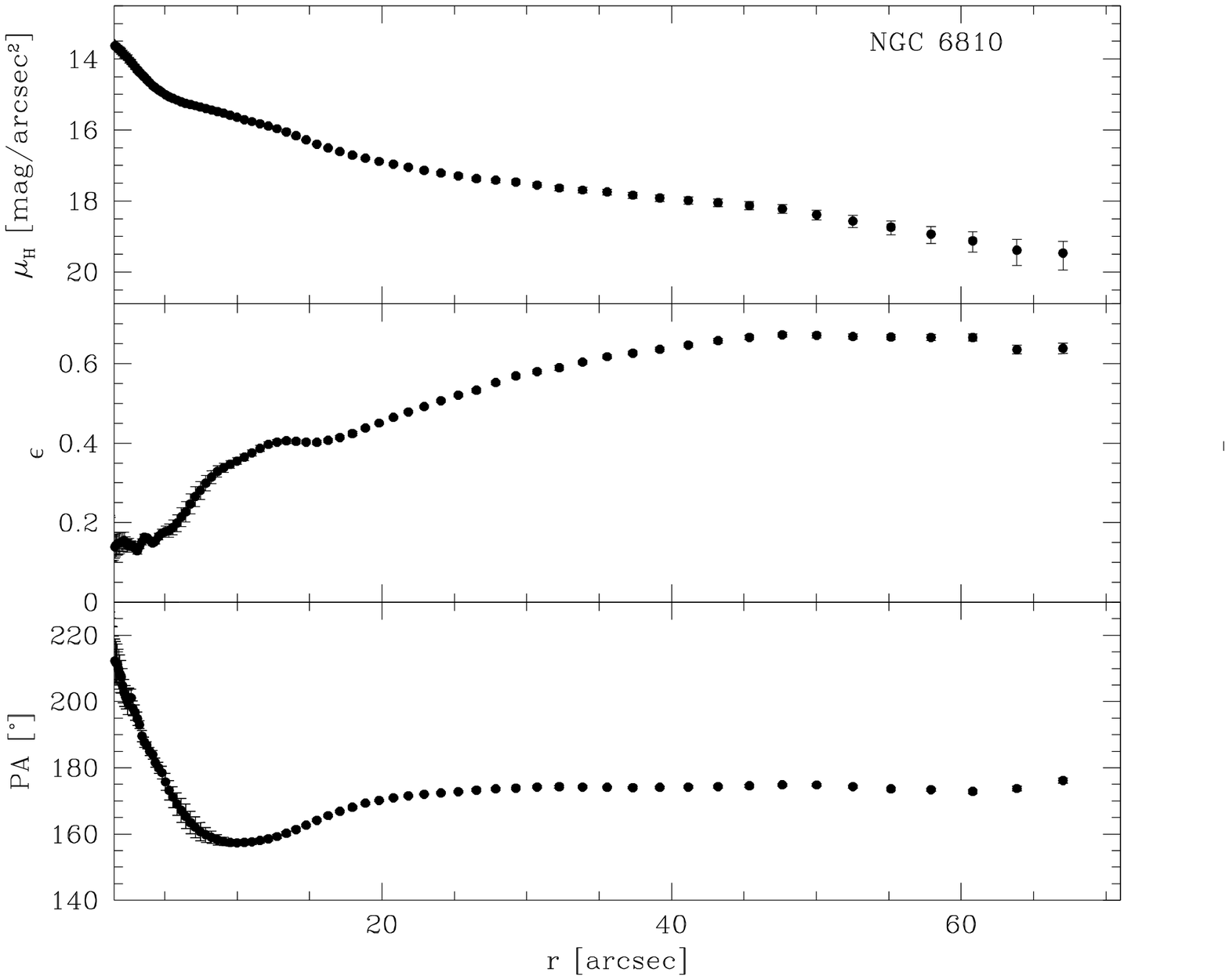}}
\caption{Surface-brightness ({\it upper panel\/}), 
  ellipticity ({\it central panel\/}) and position angle ({\it lower
  panel\/}) radial profiles for the sample galaxies. Data for NGC 949,
  NGC 2460, and NGC 6810 and in $H$ band, data for NGC 2541 are in
  $i$ band.}
\end{center}
\label{fig:profiles}
\end{figure*}

We analyzed the isophotal profiles of the background and star
subtracted images by masking the residual foreground stars, and then
fitting ellipses to the isophotes with the {\tt IRAF}\footnote{{\tt
IRAF} is distributed by NOAO, which is operated by AURA Inc., under
contract with the National Science Foundation} task {\tt ELLIPSE}.
We first allowed the centers of the ellipses to vary, to test whether
the galaxies were disturbed. Within the errors of the fits, we found
no evidence of a varying center. The ellipse fits were therefore
repeated with the ellipse centers fixed.  The resulting azimuthally
averaged surface brightness, ellipticity, and position angle profiles
for the sample galaxies are plotted in
Fig. \ref{fig:profiles}.

\section{Ionized-gas kinematics and near-infrared photometry}
\label{sec:kinematics}

The following kinematic photometric features are
noteworthy in the sample galaxies.

The central velocity gradient along the major axis is zero (NGC 949,
NGC 2541, NGC 6810) or at least less steep (NGC 2460) than that we
measured for radii larger than $\approx3''$. Further out the gas
velocity increases to reach a maximum and then remains constant out to
the last observed radius.
Non-zero gas velocities are measured along the minor axis of all the
sample galaxies, in spite of which would be expected if the gas traced
the circular velocity in the disk plane. 
The gas velocity dispersion remains low ($\la50$ \kms) except for few
radii along the major axis of NGC 2460 and in the inner regions of NGC
6810 where it reaches $\approx120$ \kms . 
Qualitatively all these data indicate that the gas is predominantly
supported by rotation, and exclude the presence of a dynamically hot
gas as found in some bulges (Bertola et al. 1995; Cinzano et
al. 1999). We attribute the minor-axis velocity gradients to the
presence of kinematically-decoupled component which is not rotating in
the plane of the stellar disk. The reversal of the gas velocities
measured along the minor axis of NGC 949 and NGC 2460, and the
receding velocities observed along the minor axis of NGC 6810 are
suggestive of an even more complex gas structure involving a warp and
an outflow, respectively.

We observed a large isophotal twist in the inner $\sim10''$ of
the sample galaxies which is associated with an increase of ellipticity
at small radii. These photometric features can be interpreted as the
signature of bulge triaxiality (e.g.  Bertola al. 1991).
The bump observed at $\sim20''$ in the surface-brightness profiles of
NGC 949 and NGC 6810 suggests the presence of a bar. There is no
signature of a bar in the photometric profiles of NGC 2450 and NGC
2541. 
For radii larger than $\sim20''$ the position angle and ellipticity
settle to almost constant values as expected when the disk component
dominates the galaxy light. The values of position angle and
ellipticity we measured at the farthest observed radius are consistent
within the errors with those in $B$ band listed in RC3, apart from the
ellipticity of NGC 6810. We attribute the different shape of the outer
isophotes of NGC 6810 to the dust lanes crossing the disk (see Panel
145, Sandage \& Bedke 1994). For all the sample galaxies the slit
position in spectroscopic observations corresponds to the major and
minor axis of the disk.

\begin{figure}[ht!]
\centering
\resizebox{\hsize}{!}{ 
\includegraphics[clip=,bb=18 200 592 490]{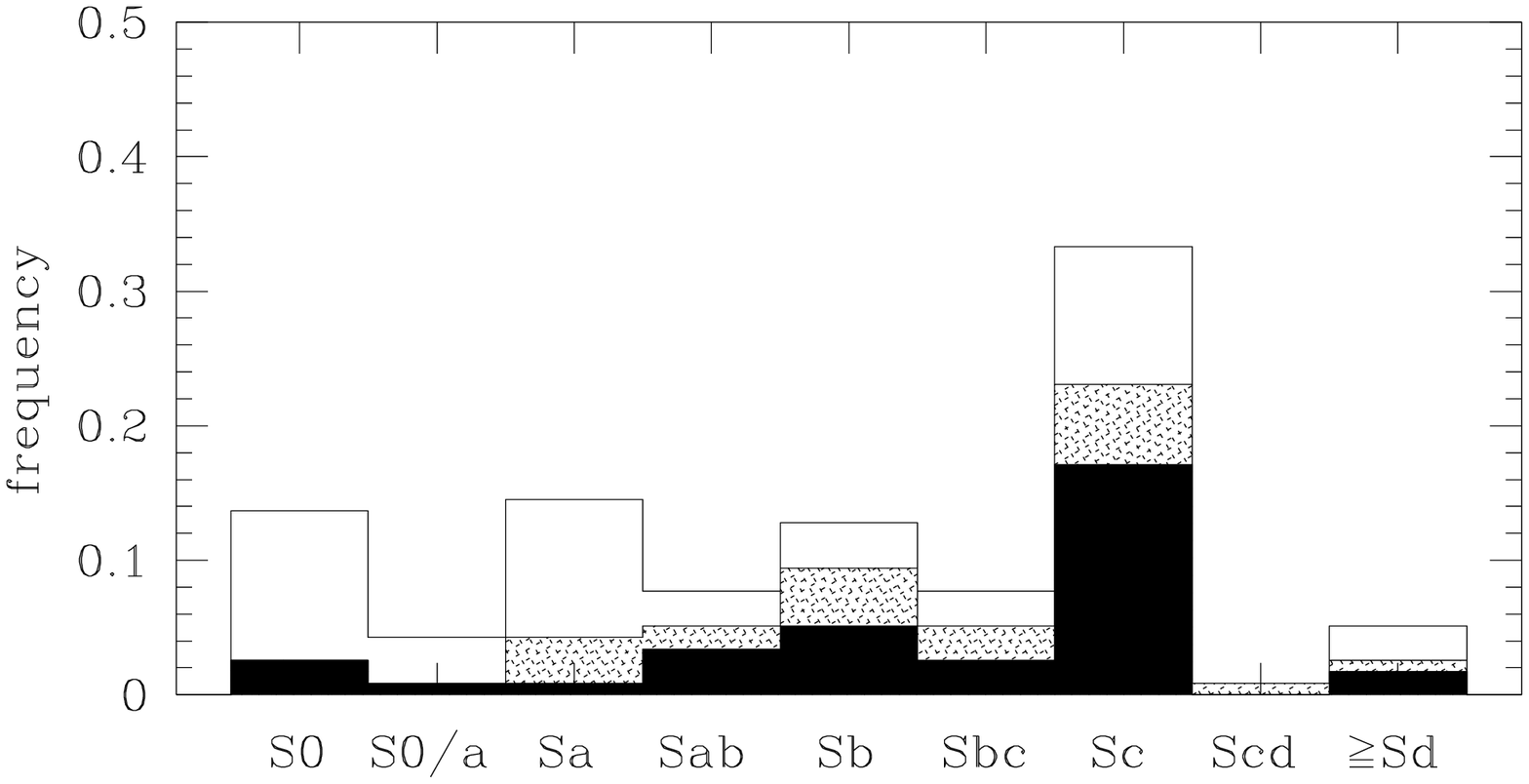}} 
\resizebox{\hsize}{!}{ 
\includegraphics[clip=,bb=18 200 592 490]{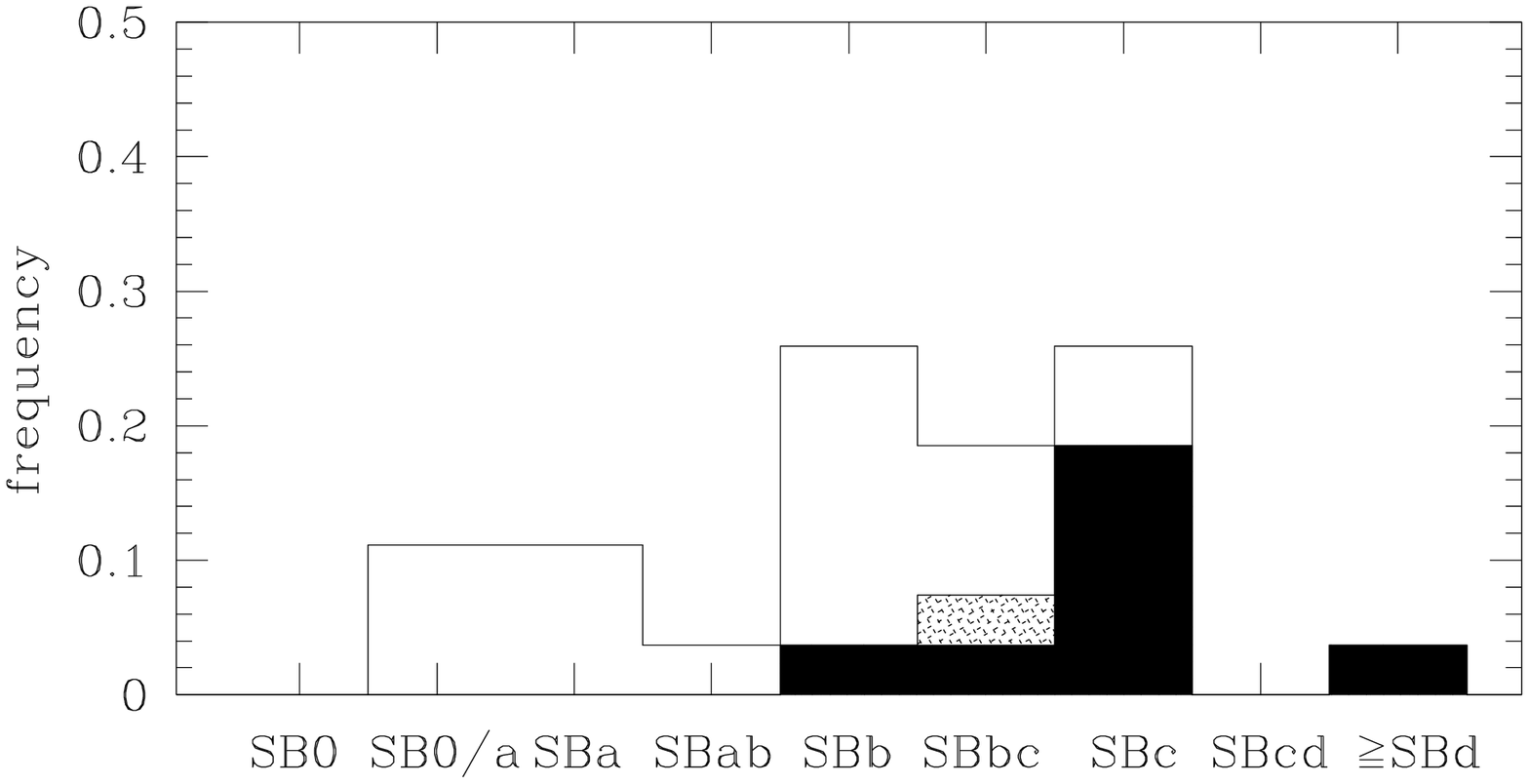}} 
\caption{Distribution of the gas velocity gradient along the optical
  minor axis of the unbarred ({\it upper panel\/}) and barred ({\it
  lower panel\/}) galaxies we collected in Table 1
  according to their RSA morphological classification. The white and
  black regions identify galaxies with (Classes O, C, and G) and
  without a minor-axis velocity gradient (Class Z), respectively. The
  shaded region identify galaxies assigned to Class I. The last bins
  of the panels include all the galaxies with a morphological type
  later than Sd and SBd, respectively.}
\label{fig:histogram}
\end{figure}

\begin{table}[!ht]
\caption{Log of the spectroscopic observations.}
\begin{small} 
\begin{tabular}{lcrcc} 
\hline
\noalign{\smallskip}
\multicolumn{1}{c}{Object}  & \multicolumn{1}{c}{Run} &  
\multicolumn{1}{c}{P.A.}  & \multicolumn{1}{c}{Axis} &
\multicolumn{1}{c}{Exp. Time} \\
 & & \multicolumn{1}{c}{[$^{\circ}$]} & & \multicolumn{1}{c}{[s]} \\
\noalign{\smallskip}
\hline 
\noalign{\smallskip} 
NGC  949 & 4 &  55 & Minor & $2\times2700$ \\
NGC  949 & 1 & 145 & Major & $2\times2700$ \\
NGC 2460 & 4 & 130 & Minor & $2\times2700$ \\
NGC 2460 & 1 &  40 & Major & $2\times2700$ \\
NGC 2541 & 4 &  75 & Minor & $2\times2700$ \\
NGC 2541 & 1 & 165 & Major & $1\times2700$ \\
NGC 6810 & 3 &  86 & Minor & $1\times1800$ \\
NGC 6810 & 2 & 176 & Major & $1\times2400$ \\
\noalign{\smallskip}  
\hline 
\noalign{\smallskip}  
\end{tabular}  
\end{small}
\label{tab:log}
\end{table}

\section{Discussion and conclusions}
\label{sec:conclusions}

By combining the new kinematic data of our sample galaxies with those
available in the literature for the RSA galaxies we have the minor-axis
velocity profiles of the ionized-gas component for a sample of
142 disk galaxies. In Fig. \ref{fig:histogram} we show the
distribution of the gas velocity gradients measured along the optical
minor axis these galaxies according to their RSA morphological
classification.
Although the effective frequency of the presence gas velocity
gradients along the minor axis of bright disk galaxies can be properly
addressed only with the analysis of a magnitude limited sample,
Fig. \ref{fig:histogram} favors a picture in which this is a
widespread phenomenon which is not limited to few peculiar objects:
95/142 ($67\%$) of the studied galaxies shows motions
(including irregular ones) along the minor axis.

The presence of a minor-axis velocity gradient in 16/25 ($64\%$) of
the available barred galaxies is explained as due to the non-circular
(e.g. Athanassoula 1992) and off-plane (e.g. Friedli \& Benz 1993)
motions induced on the gaseous component by the tumbling triaxial
potential of the bar.

We find that 54/117 ($46\%$) of the unbarred galaxies shows a
minor-axis gas velocity gradient and the phenomenon is particularly
frequent in the earliest morphological types. If we consider also the
unbarred galaxies with an irregular velocity profile along the
minor axis, the fraction rises to 75/117 ($64\%$).
Most of these velocity gradients has to be attributed to the presence
of non-detected bars since $\approx 40\%$ of the spiral galaxies
without a strong optical bar reveal a strong bar when observed in near
infrared (Eskridge et al. 2000). However, it must be noted that in
addition to the presence of a bar other mechanisms can be invoked to
explain the observed velocity gradients.

A large fraction of the S0 galaxies (6/16, $38\%$) listed in Table
1 is characterized by an overall gas velocity profile
along the optical minor axis. If the gas in all S0 galaxies is of
external origin (Bertola, Buson \& Zeilinger 1992; Kuijken, Fisher \&
Merrifield 1996) then the observed kinematics can be straightforwardly
interpreted as due to accreted gas which is rotating in a
warped/inclined disk with respect to the stellar disk (e.g. NGC 4753,
Steinman-Cameron, Kormendy \& Durisen 1992). This is a stable
configuration in a barred potential when gas is settled onto the
so-called anomalous orbits (e.g. NGC 128, Emsellem \& Arsenault 1996).

The presence of non-zero velocities confined in the nuclear regions
along the minor axis is particularly frequent in bulge-dominated
spirals. Since there is no significant trend in the bar fraction as a
function of morphology in either the optical or near infrared
photometry (Eskridge et al. 2000), this finding suggests that at least
some of the minor-axis velocity gradients are closely related to bulge
prominence.
Indeed the intrinsic shape of bulges is triaxial (Bertola et al. 1991)
and two equilibrium planes are allowed for the gaseous component. We
are left with two alternative scenarios as viable mechanisms to
explain minor-axis velocity gradients in early-type spirals.
All the gaseous component lies on the principal plane perpendicular
to the bulge short axis and moves onto closed elliptical orbits, which
become nearly circular at large radii (de Zeeuw \& Franx 1989; Gerhard
et al. 1989; Corsini et al. 2003). A central velocity gradient is
measured along both the major and minor axis as due to the orientation of
the inner elliptical orbits lying on the disk plane and seen at
an intermediate angle between their intrinsic major and minor
axes. The Minor-axis gas velocity profile drops to zero where
elliptical orbits become circular. Alternatively, only the outer
gaseous component lies on the plane perpendicular to the short axis of
bulge, while the inner gas is rotating in the principal plane
perpendicular to the long axis of the bulge giving rise to an inner polar
disk (Corsini et al. 2003). If this is the case the central velocity
gradient observed along the disk minor axis is associated with a central
zero-velocity plateau (or at least to a shallower velocity gradient)
along the disk major axis.
 
\begin{acknowledgements}
We thank the referee, O. Sil'chenko, for the suggestions which helped
us to improve the paper. We are also grateful to J. A. L. Aguerri for
useful discussions. This research has made use of the NASA/IPAC
Infrared Science Archive, which is operated by the Jet Propulsion
Laboratory, California Institute of Technology, under contract with
the National Aeronautics and Space Administration.
\end{acknowledgements}

\clearpage
\newpage
\addtocounter{table}{-4}
\onecolumn
\setlongtables
\begin{tiny}
\begin{landscape}
\begin{longtable}{l l l c c r r r c r r l}
\caption{Ionized-gas velocity gradients observed along the minor axis of the 
  disk galaxies listed in RSA Catalog.}\\
\endfirsthead
\multicolumn{12}{c}{Tab. 1: {\tiny (continue).}} \\
\noalign{\smallskip}
\noalign{\smallskip}
\noalign{\smallskip}
\hline
\noalign{\smallskip}
\endhead
\noalign{\smallskip}
\hline
\endfoot
\hline
\endlastfoot
\hline
\noalign{\smallskip}
\multicolumn{1}{l}{Object} &
\multicolumn{2}{c}{Morph. Type} &
\multicolumn{1}{c}{$D_{25}$} &
\multicolumn{1}{c}{$i$} &
\multicolumn{1}{c}{PA$_{\rm MJ}$} &
\multicolumn{1}{c}{$B^0_T$} &
\multicolumn{1}{c}{$cz$} &
\multicolumn{1}{c}{Class} &
\multicolumn{1}{c}{Obs. PA$_{\rm MJ}$} &
\multicolumn{1}{c}{Obs. PA$_{\rm MN}$} &
\multicolumn{1}{l}{References} \\
\noalign{\smallskip}
\multicolumn{1}{l}{} &
\multicolumn{1}{l}{[RSA]}  &
\multicolumn{1}{l}{[RC3]} &
\multicolumn{1}{c}{[$'$]}  &
\multicolumn{1}{c}{[$^\circ$]} &
\multicolumn{1}{c}{[$^\circ$]} &
\multicolumn{1}{c}{[mag]} &
\multicolumn{1}{c}{[\kms]} &
\multicolumn{1}{c}{} &
\multicolumn{1}{c}{[$^\circ$]} &
\multicolumn{1}{c}{[$^\circ$]} &
\multicolumn{1}{l}{} \\
\noalign{\smallskip}
\multicolumn{1}{l}{(1)} &
\multicolumn{1}{l}{(2)}  &
\multicolumn{1}{l}{(3)} &
\multicolumn{1}{c}{(4)}  &
\multicolumn{1}{c}{(5)} &
\multicolumn{1}{c}{(6)} &
\multicolumn{1}{c}{(7)} &
\multicolumn{1}{c}{(8)} &
\multicolumn{1}{c}{(9)} &
\multicolumn{1}{c}{(10)} &
\multicolumn{1}{c}{(11)} &
\multicolumn{1}{l}{(12)} \\
\noalign{\smallskip}
\hline
\noalign{\smallskip}
\multicolumn{12}{l}{S0} \\
\object{IC 4889	}&S0$_{1/2}$(5)	&E5$+$		&   2.9 & 56   	&0	&11.91	&2526	&O &     0&  90	& Corsini et al. 2000	\\
\object{NGC 1947}&S0$_3$(0)	&S0 pec		&   3.0 & 32   	&119	&11.50	&1157	&O &   120&  30	& Mollenhoff 1982	\\
		 &		&		&       &      	&	&	&	&  &   120&  30	& Bertola et al. 1992b	\\
\object{NGC 2685}&S0$_3$(7) pec	&(R)SB0$^+$ pec	&   4.5 & 60   	&38	&11.82	&869	&G &    40& 130	& Schechter \& Gunn. 1978\\
\object{NGC 3414}&S0$_1$	&S0 pec 	&   3.5 & 45   	&179$^*$&11.86	&1434	&G &    25& 115	& Lake \& Dressler 1986	\\
\object{NGC 3998}&S0$_1$	&S0$^0$(r)? 	&   2.7 & 34   	&140	&11.49	&1066	&C &\ldots&  45	& Blackman et al. 1983	\\
		 &		&		&       &      	&	&	&	&  &   140&  50	& Fisher 1997		\\
\object{NGC 4026}&S0$_{1/2}$(9)	&S0 sp		&   5.2 & 80   	&178	&11.59	&930	&G &   178&  88	& Fisher 1997		\\
\object{NGC 4111}&S0$_1$(9)	&S0$^+$(r): sp  &   4.6 & 83   	&150	&11.60	&815	&G &   150&  60	& Fisher 1997		\\
\object{NGC 4383}&S0:		&Sa pec?	&   2.0 & 60   	&28	&12.38	&1663	&Z &   208& 118	& Rubin et al. 1999	\\
\object{NGC 4710}&S0$_3$(9)	&S0$^0$(r)? /	&   4.9 & 81   	&27	&11.80	&1119	&G &    27& 117	& Rubin et al. 1999	\\
\object{NGC 5084}&S0$_1$(8)	&S0 sp		&   9.3 & 88   	&80	&11.28	&1706	&G &    80& 170	& Zeilinger et al. 1990	\\
\object{NGC 5128}&S0+S pec	&S0 pec		&  25.7 & 40   	&35	&7.30	&538	&O &    35& 125	& van Gorkom et al. 1990\\
\object{NGC 5266}&S0$_3$(5) pec (prolate)&S0$^-$:&  3.2 & 50   	&19	&11.50	&3074	&O &   114&  15	& Caldwell 1984		\\
\object{NGC 5363}& [S0$_3$(5)]  &I0             &   4.1 & \ldots&135    &\ldots &1121   &Z &   135&  45 & Sharples et al. 1983  \\
\object{NGC 5866}&S0$_3$(8)	&S0$^+$ sp	&   4.7 & 68   	&128	&10.83	&769	&Z &   128&  38	& Fisher 1997		\\
\object{NGC 5898}&S0$_{2/3}$(8)	&E0		&   2.2 & 25   	&37$^*$	&11.92	&2209	&O &    66& 156	& Caon et al. 2000	\\
\object{NGC 7332}&S0$_{2/3}$(0)	&S0 pec sp	&   4.1 & 78   	&155	&11.93	&1202	&O &   155&  65	& Fisher at al. 1994	\\
\noalign{\smallskip}
\multicolumn{12}{l}{S0a/Sa,SB0/Sa,SB0/SBa}\\			                 
\object{IC 5063	}&S0$_3$(3) pec/Sa&S0$^+$(s):	&   2.1 & 49   	&116	&12.58	&3408	&G &   115&  25	& Danziger et al. 1981	\\
\object{NGC 3489}&S03/Sa	&SAB0(rs)  	&   3.5 & 56   	&70	&11.15	&690	&Z &	76& 166	& Caon et al. 2000	\\
\object{NGC 2787}&SB0/a		&SB0$^+$(r)     &   3.2 & 51   	&117	&11.61	&649	&O &   117&  27	& Vega Beltr\`an, priv. comm.\\
\object{NGC 3941}&SB0$_{1/2}$/a	&SB0$^0$(s)	&   3.5 & 50   	&10	&11.25	&934	&O &    10& 100	& Fisher 1997		\\
\object{NGC 4036}&S0$_3$(8)/Sa	&S0$^-$		&   4.3 & 69   	&85	&11.49	&1397	&O &    85& 175	& Fisher 1997		\\
\object{NGC 4546}&SB0$_1$/Sa	&SB0$^-$(s):  	&   3.3 & 66   	&78	&11.25	&1014	&O &   135&  45	& Bettoni \& Galletta 1997\\
\object{NGC 7049}&S0$_3$(4)/a	&S0$^0$(s)  	&   4.3 & 47   	&57	&11.57	&2198	&C &    57& 147	& Corsini et al. 2003	\\
\object{NGC 7377}&S0$_1$(3)/Sa	&S0$^+$(s)  	&   3.0 & 34   	&101	&11.93	&3339	&O &   101&  11	& Corsini et al. 2003	\\
\noalign{\smallskip}
\multicolumn{12}{l}{Sa}\\					                 	 	 
\object{NGC 1316}&Sa pec (merger)&SAB0$^0$(s) pec& 12.0 & 46   	&50$^*$	&9.40	&1793	&I &    50& 140	& Vega Beltr\`an, priv. comm.\\
\object{NGC 2782}&Sa(s) pec	&SABa(rs) pec	&   3.5 & 43   	&\ldots	&12.01	&2532	&C &    90&   0	& Sakka et al. 1973	\\
		 &		&		&       &      	&	&	&	&C &    75& 165	& Jogee et al. 1999	\\
\object{NGC 2855}&S(r)		&(R)S0/a(rs)	&   2.5 & 27   	&130$^*$&12.29	&1910	&C &   120&  30	& Corsini et al. 2002	\\
\object{NGC 2992}&Sa (tides)	&Sa pec		&   3.5 & 75   	&15$^*$	&12.16	&2334	&I &    17& 100	& M\'arquez et al. 1998	\\
\object{NGC 3626}&Sa		&(R)S0$^+$(rs)	&   2.7 & 45   	&157	&11.62	&1438	&I &   157&  67	& Haynes et al. 2000	\\
\object{NGC 3885}&Sa		&S0/a(s) 	&   2.4 & 69   	&123	&12.30	&1918	&O &   123&  33	& Corsini et al. 2003	\\
\object{NGC 3900}&Sa(r)		&S0$^+$(r)	&   3.2 & 59   	&2	&12.17	&1702	&C &     2&  92	& Haynes et al. 2000	\\
\object{NGC 4224}&Sa		&Sa(s): sp	&   2.6 & 69   	&57	&12.52	&2574	&C &    57& 147	& Corsini et al. 2003	\\
\object{NGC 4235}&Sa		&Sa(s) sp	&   4.2 & 82   	&48	&11.89	&2343	&I &    48& 138	& Corsini et al. 2003	\\
\object{NGC 4424}&S(a?) pec	&SBa(s) sp	&   3.6 & 62   	&95	&11.94	&447	&G &    95& 185	& Kenney et al. 1996	\\
\object{NGC 4586}&Sa		&Sa(s): sp	&   4.0 & 73   	&115	&12.11	&813	&C &   115&  25	& Corsini et al. 2003	\\
\object{NGC 4698}&Sa		&Sab		&   4.0 & 53   	&170	&11.24	&1032	&C &   170&  80	& Sarzi et al. 2000	\\
\object{NGC 4772}&Sa:		&Sa		&   3.4 & 62   	&147	&11.89	&1042	&C &   147&  57	& Haynes et al. 2000	\\
\object{NGC 4845}&Sa		&Sab sp	        &   5.0 & 79   	&89	&11.42	&1125	&G &    78& 168	& Bertola et al. 1989	\\
\object{NGC 4984}&Sa(s)		&(R)SAB0$^+$(rs)&   2.8 & 38   	&90	&12.03	&1243	&O &    90&   0	& Corsini et al. 2003	\\
\object{NGC 5854}&Sa		&SB0$^+$(s) sp	&   2.8 & 77   	&55	&12.69	&1668	&Z &    55& 145	& Haynes et al. 2000	\\
\object{NGC 7213}&Sa(rs)	&Sa(s): sp	&   3.1 & 27   	&124$^*$&11.13	&1803	&O &   124&  34	& Corsini et al. 2003	\\
\noalign{\smallskip}
\multicolumn{12}{l}{SBa}\\
\object{NGC 2217}&SBa(s)	&(R)S0		&   4.5 & 21   	&34$^*$	&11.35	&1609	&C &   110&  20	& Bettoni et al. 1990	\\
\object{NGC 3081}&SBa(s)	&(R)SAB(r)$^0$	&   2.1 & 40   	&158	&12.59	&2391	&O &\ldots&  74	& Storchi-Bergmann et al. 1996\\
\object{NGC 7079}&SBa		&SB0(s)$^0$	&   2.1 & 53   	&82	&12.36	&2663	&G &    82& 170	& Bettoni \& Galletta 1997\\
\noalign{\smallskip}
\multicolumn{12}{l}{Sab}\\
\object{NGC 2460}& Sab(s)       & Sa(s)         &   2.5 & 42    &40     &12.29  &1442   &O &    40& 130 & this paper\\
\object{NGC 4138}&Sab(r)	&S0$^+$(r)	&   2.6 & 50   	&150	&12.10	&1039	&C &   150&  60	& Jore et al. 1996\\
\object{NGC 4151}&Sab		&(R$'$)SABab(rs):&  6.3 & 46   	&50	&10.71	&995$^*$&Z &    48& 138	& Winge et al. 1999	\\
\object{NGC 4388}&Sab		&Sb(s): sp	&   5.6 & 83   	&92	&10.79	&2538	&I &    90&   0	& Rubin et al. 1999	\\
\object{NGC 4450}&Sab pec	&Sab(s)	        &   5.2 & 43   	&175	&10.75	&2048	&I &   175&  85	& Rubin et al. 1999	\\
\object{NGC 4569}&Sab(s)I-II	&SABab(rs)	&   9.6 & 65   	&23	&9.79	&$-311$	&Z &  23.5&113.5& Rubin et al. 1999	\\
\object{NGC 4579}&Sab(s)II	&SABb(rs)	&   5.9 & 38   	&95	&10.29	&1627	&O &   275& 185	& Rubin et al. 1999	\\
\object{NGC 4826}&Sab(s)II	&(R)Sab(rs)	&  10.0 & 59   	&115	&8.82	&474	&Z &   115&  25	& Rubin 1994		\\
\object{NGC 7177}&Sab(r)II.2	&SABab(r)	&   3.1 & 51   	&90	&11.47	&1112	&Z &    90&   0	& M\'arquez et al. 2002	\\
\noalign{\smallskip}
\multicolumn{12}{l}{SBab}\\
\object{NGC 4419}&SBab:		&SBa(s)		&   3.3 & 74   	&133	&11.60	&$-224$	&O &   133&  43	& Rubin et al. 1999	\\
\noalign{\smallskip}
\multicolumn{12}{l}{Sb}\\
\object{NGC  224}&SbI-II	&Sb(s)		& 190.5 & 72   	&35	&3.36	&$-295$	&I &\ldots& 128	& Sandqvist et al. 1989 \\
		 &		&		&       &      	&	&	&	&  &    38& 128	& Pellet 1976	        \\
		 &		&		&       &      	&	&	&	&  &    38& 128	& Rubin \& Ford 1970	\\
\object{NGC  891}&Sb on edge	&Sb(s)? sp	&  13.5 & 81   	&22	&9.37	&528$^*$&Z &    22& 112	& Keppel et al. 1991	\\
\object{NGC 1961}&Sb(rs)II pec	&SABc(rs)	&   4.6 & 50   	&85	&11.01	&3983	&Z &    85& 175	& Rubin et al. 1979	\\
\object{NGC 2841}&Sb	&Sb(r)	&   8.1 & 64   	&147	&9.58	&635	&G &    150& 62	& Sil'chenko et al. 1997	\\
\object{NGC 3031}&Sb(r)I-II	&Sab(s)	        &  26.9 & 59   	&157	&7.39	&$-49$	&G &   150&  60	& Goad 1974		\\
\object{NGC 3504}&Sb(s)/SBb(s)I-II&(R)SABab(s)	&   2.7 & 39   	&57$^*$	&11.65	&1518	&Z &   152&  62	& Burbidge, et al. 1960	\\
\object{NGC 3521}&Sb(s)II-III	&SABbc(rs)	&  11.0 & 63   	&163	&9.29	&782	&Z &   163&  73	& Zeilinger et al. 2001	\\
\object{NGC 4192}&SBII:		&SABb(s)	&   9.8 & 75   	&155	&10.02	&$-126$	&I &   155&  65	& Rubin et al. 1999	\\
\object{NGC 4258}&Sb(s)II	&SABbc(s)	&  18.6 & 68   	&150	&8.53	&480	&G &   148&  58	& Chincarini \& Walker 1967\\
		 &		&		&       &      	&	&	&	&  &   150&\ldots& Sofue et  al. 1999	\\
\object{NGC 4527}&RSb(rs)II	&SABbc(s)	&   6.2 & 71   	&67	&10.66	&1727	&I &    67& 157	& Rubin et al. 1999	\\
\object{NGC 5172}&SbI		&SABbc(rs):	&   3.3 & 59   	&103	&12.10	&4337	&Z &   103&  13	& M\'arquez et al. 2002	\\
\object{NGC 6810}& Sb           & Sab(s):sp     &   3.3 & 78    &176    &11.49  &1958   &I &   176&  86 & this paper            \\    
\object{NGC 6951}&Sb/SBb(rs)I.3	&SABbc(rs)	&   3.9 & 34   	&170	&10.71	&1331	&I &   138&  48	& Perez et al. 2000	\\
\object{NGC 7217}&Sb(r)II-III	&(R)Sab(r)	&   3.9 & 34   	&95	&10.53	&935	&G &    86&  10	& Peterson et al. 1978	\\
\object{NGC 7331}&Sb(rs)I-II	&Sb(s)		&  10.5 & 70   	&171	&9.38	&835	&Z &   171&  77	& Rubin et al. 1965	\\
\noalign{\smallskip}
\multicolumn{12}{l}{SBb}\\
\object{NGC 1300}&SBb(s)I.2	&SBbc(rs)	&   6.2 & 49   	&106	&10.77	&1592	&C &   103& 185	& Peterson et al. 1980	\\
\object{NGC 1365}&SBb(s)I	&SBb(s)		&  11.2 & 57   	&32	&9.93	&1675	&G &   225& 318	& Lindblad et al. 1996	\\
\object{NGC 3992}&SBb(rs)I	&SBbc(rs)	&   7.6 & 52   	&68	&10.26	&1059	&O &    67& 152	& Wilke et al. 2000	\\
\object{NGC 4639}&SBb(r)II	&SABbc(rs)	&   2.8 & 48   	&123	&11.85	&898	&C &   124&  34	& Rubin et al. 1999	\\
\object{NGC 5728}&SBb(rs)II	&SABa(r):	&   3.1 & 55   	&30	&11.65	&2885	&C &    33& 123	& Rubin et al. 1980	\\
\object{NGC 5850}&SBb(sr)I-II	&SBb(r)		&   4.3 & 30   	&140	&11.39	&2483	&Z &   163&  73	& Higdon et al. 1980	\\
\object{NGC 6300}&SBb(s)II pec	&SBb(rs)	&   4.5 & 49   	&118	&10.20	&1064	&C &   111&  21	& Buta 1987		\\
\object{NGC 6951}&Sb/SBb(rs)I.3	&SABbc(rs)	&   3.9 & 34   	&170	&10.71	&1331	&I &   138&  48	& Perez et al. 2000	\\
\noalign{\smallskip}
\multicolumn{12}{l}{Sbc}\\
\object{NGC 3310}&Sbc(r) pec	&SABbc(r) pec	&   3.1 & 40   	&\ldots	&10.95	&1018	&I &   150&  60 & Mulder \& van Driel 1996\\
\object{NGC 4433}&SbcIII	&SABab(s)	&   2.2 & 65   	&5	&12.64	&2942	&G &     0&  90	& Chromey 1974		\\
\object{NGC 4501}&Sbc(s)II	&Sb(rs)	        &   6.9 & 59   	&140	&9.86	&2120	&Z &   320& 210	& Rubin et al. 1999	\\
\object{NGC 4580}&Sbc(r)II	&SABs(rs) pec	&   2.1 & 40   	&165	&12.49	&1227	&G &   345& 255	& Rubin et al. 1999	\\
\object{NGC 5194}&Sbc(s)I-II	&SAbc(s) pec	&  11.2 & 53   	&163	&8.67	&463	&I &   170&  80	& Kuno et al. 1997	\\
\object{NGC 5248}&Sbc(s)I-II	&SABbc(rs)	&   6.2 & 44   	&110	&10.63	&1189	&Z &   113&  23	& Burbidge et al. 1962	\\
\object{NGC 6221}&Sbc(s)II-III	&SBc(s)		&   3.5 & 47   	&5	&9.77	&1350	&I &     5&  95	& Vega Beltr\`an et al. 1998\\
\object{NGC 6574}&Sbc(r)II	&SABbc(rs):	&   1.4 & 40   	&160	&11.79	&2321	&C &   159&  66	& Demoulin 1969		\\
\object{NGC 6814}&Sbc(rs)I-II	&SABbc(rs)	&   3.0 & 21   	&89$^*$	&11.32	&1509	&Z &    90&   0	& Schulz et al. 1994	\\
\noalign{\smallskip}
\multicolumn{12}{l}{SBbc}\\
\object{NGC 2336}&SBbc(r)I	&SABbc(r)	&   7.1 & 58   	&178	&10.61	&2205	&C &     5&  94	& Wilke et al. 1999	\\
\object{NGC 7479}&SBbc(s)I-II	&SBc(s)		&   4.1 & 41   	&25	&11.22	&2394	&I &    35& 125	& Wilke et al. 2000	\\
\noalign{\smallskip}
\multicolumn{12}{l}{Sc}\\
\object{NGC  157}&Sc(s)II-III	&SABbc(rs)	&   4.2 & 50   	&30	&10.59	&1730	&Z &\ldots& 130	& Ryder et al. 1998	\\
		 &		&		&       &      	&	&	&	&  &   224&\ldots& Fridman et al. 2001	\\
\object{NGC  253}&Sc(s)		&SABc(s)	&  27.5 & 77   	&52	&7.09	&240	&I &    52& 321 & Prada et al. 1998	\\
\object{NGC  949}& Sc(s)III     &Sb(rs):?       &   2.4 & 58    &145    &11.78  &596    &I &   145&  55 & this paper            \\
\object{NGC 1084}&Sc(s)II.2	&Sc(s)	        &   3.2 & 56   	&35	&10.90	&1414	&Z &    33& 123 & Burbidge et al. 1963	\\
\object{NGC 1087}&Sc(s)III.3	&SABc(rs)	&   3.7 & 53   	&5	&10.97	&1508	&Z &     0&  90 & Blackman 1980a	\\
\object{NGC 1792}&Sc(s)II	&SAbc(rs)	&   5.2 & 61   	&137	&10.36	&1218	&Z &   140&  50 & Rubin et al. 1964	\\
\object{NGC 2541}&Sc(s)III      & Scd(s)        &   6.3 & 60    &165    &11.57  &628    &C &   165&  75 & this paper            \\
\object{NGC 2748}&Sc(s)II-III	&SAbc		&   3.0 & 68   	&38	&11.59	&1456	&G &    44& 134 & Reshetnikov \& Combes 1994\\
\object{NGC 2776}&Sc(s)I	&SABc(rs)	&   3.0 & 27   	&93$^*$	&11.96	&2618	&G &    90&   0 & Carozzi 1979		\\
\object{NGC 2903}&ScIII pec	&SABbc(rs)	&  12.6 & 62   	&17	&9.11	&565	&O &     9& 103 & Walker 1989		\\
\object{NGC 2998}&Sc(s)II	&SABc(rs)	&   2.9 & 63   	&53	&12.52	&4767	&Z &    53& 143 & Rubin et al. 1980	\\
\object{NGC 3079}&Sc pec	&SBc(s) sp	&   7.9 & 81   	&165	&10.41	&1101	&Z &   169&  79	& Veilleux et al. 1999	\\
\object{NGC 3495}&Sc(s)III	&Sd:		&   4.9 & 77   	&20	&10.74	&1086	&Z &    20& 110	& Rubin et al. 1980	\\
\object{NGC 3672}&Sc(s)I-II	&Sc(s)		&   4.2 & 63   	&12	&11.41	&1857	&C &     8&  98	& Rubin et al. 1977	\\
\object{NGC 3684}&Sc(s)II	&Sbc(rs)	&   3.1 & 47   	&130	&11.82	&1394	&Z &   128&  38	& Blackman 1980b	\\
\object{NGC 4100}&Sc(s)I-II	&(R$'$)Sbc(rs)	&   5.4 & 72   	&167	&11.05	&1131	&G &\ldots&  78	& Afanaz'ev et al. 1992	\\
\object{NGC 4212}&Sc(s)II-III	&Sc:		&   3.2 & 52   	&75	&11.35	&377	&Z &    75& 165	& Rubin et al. 1999	\\
\object{NGC 4237}&Sc(r)II.8	&SABbc(rs)	&   2.1 & 50   	&108	&12.18	&865	&Z &   108&  18	& Rubin et al. 1999     \\
\object{NGC 4298}&Sc(s)III	&Sc(rs)	        &   3.2 & 56   	&140	&11.62	&1122	&I &   140&  50	& Rubin et al. 1999	\\
\object{NGC 4303}&Sc(s)I.2	&SABbc(rs)	&   6.5 & 27   	&162$^*$& 10.12	&1607	&I &   318& 228	& Rubin et al. 1999	\\
\object{NGC 4321}&Sc(s)I	&SABbc(s)	&   7.4 & 32   	&30	&9.98	&1579	&O &   350& 260	& Rubin et al. 1999	\\
\object{NGC 4420}&Sc(s)III	&SBbc(r):	&   2.0 & 62   	&8	&12.28	&1711	&Z &     8&  98	& Rubin et al. 1999	\\
\object{NGC 4536}&Sc(s)I	&SABbc(rs)	&   7.6 & 65   	&130	&10.58	&1894	&Z &\ldots&  40	& Afanaz'ev et al. 1992	\\
\object{NGC 4567}&Sc(s)II-III	&Sbc(rs)	&   3.0 & 48   	&85	&11.79	&2213	&I &    85& 175	& Rubin et al. 1999	\\
\object{NGC 4568}&Sc(s)II-III	&Sbc(rs)	&   4.6 & 65   	&23	&11.18	&2260	&Z &    23& 113	& Rubin et al. 1999	\\
\object{NGC 4571}&Sc(s)II-III	&Sd(r)		&   3.6 & 27   	&55	&11.73	&284	&Z &    40& 130	& Rubin et al. 1999	\\
\object{NGC 4595}&Sc(s)II.8	&SABb(rs)?	&   1.7 & 50   	&110	&12.62	&589	&Z &   110& 200	& Rubin et al. 1999	\\
\object{NGC 4647}&Sc(s)III	&SABc(rs)	&   2.9 & 38   	&125	&11.81	&1421	&O &   124& 214	& Rubin et al. 1999	\\
\object{NGC 4651}&Sc(r)I-II	&Sc(rs)	        &   4.0 & 49   	&80	&11.04	&788	&C &    71& 161	& Rubin et al. 1999	\\
\object{NGC 4808}&Sc(s)III	&Scd(s):	&   2.8 & 66   	&127	&11.82	&738	&Z &   127&  37	& Rubin et al. 1999	\\
\object{NGC 4945}&Sc		&SBcd(s): sp	&  20.0 & 81   	&43	&7.43	&581	&G &    42& 132	& Peterson 1980		\\
\object{NGC 5012}&Sc(rs)I-II	&SABc(rs)	&   2.9 & 55   	&10	&12.49	&2661	&Z &    10& 170	& M\'arquez et al. 2002	\\
\object{NGC 5457}&Sc(s)I	&SABcd(rs)	&  28.8 & 21  	&\ldots	&8.21	&221	&Z &    35& 125	& Comte et al. 1979	\\
\object{NGC 5480}&Sc(s)III	&Sc(s):	        &   1.7 & 49   	&0	&12.49	&1823	&I &   177&  87	& M\'arquez et al. 2002	\\
\object{NGC 5962}&Sc(r)II.3	&Sc(r)		&   3.0 & 45   	&110	&11.73	&1993	&C &   110&  20	& M\'arquez \& Moles 1996\\
\object{NGC 6070}&Sc(s)I	&Scd(s)	        &   3.5 & 59   	&62	&11.58	&2007	&I &    62& 152	& M\'arquez et al. 2002	\\
\object{NGC 6106}&Sc(rs)II.3	&Sc		&   2.5 & 57   	&140	&12.22	&1464	&Z &   140&  50	& M\'arquez et al. 2002	\\
\object{NGC 6946}&Sc(s)II	&SABcd(rs)	&  11.5 & 32   	&\ldots	&7.78	&7	&O &    67& 157	& Munoz Tunon et al. 1987\\
\object{NGC 7541}&Sc(s)II	&SBbc(rs): pec	&   3.5 & 70   	&102	&11.57	&2669	&Z &\ldots&  15	& Kyazumov 1980		\\
\noalign{\smallskip}
\multicolumn{12}{l}{SBc}\\
\object{NGC 4064}&SBc(s):	&SBa(s): pec	&   4.4 & 67   	&150	&11.75	&1000	&G &   150&  60	& Rubin et al. 1999\\
\object{NGC 4178}&SBc(s)II	&SBdm(rs)	&   5.1 & 70   	&30	&11.35	&240	&Z &    30& 120	& Rubin et al. 1999	\\
\object{NGC 4294}&SBc(s)II-III	&SBcd(s)	&   3.2 & 68   	&155	&11.83	&421	&Z &   155&  65	& Rubin et al. 1999	\\
\object{NGC 4535}&SBc(s)I.3	&SABc(s)	&   7.1 & 45   	&0	&10.32	&1973	&C &     0&  90	& Rubin et al. 1999	\\
\object{NGC 4654}&SBc(rs)II	&SABcd(rs)	&   4.9 & 55   	&128	&10.75	&1035	&Z &   128&  38	& Rubin et al. 1999	\\
\object{NGC 4713}&Sbc(s)II-III	&SABd(rs)	&   2.7 & 51   	&100	&11.85	&631	&Z &   100&  10	& Rubin et al. 1999	\\
\object{NGC 7070}&SBc(s)II.8	&Scd(s)	        &   2.3 & 39    &23$^*$ &12.55	&2369$^*$&Z&    29& 119	& Sharples at al. 1983  \\
\noalign{\smallskip}
\multicolumn{12}{l}{Scd}\\
\object{NGC 4490}&Scd III pec	&SBd(s) pec	&   6.3 & 61  	&125	&9.81	&594	&I &   125&  62	& Duval 1981		\\
\noalign{\smallskip}
\multicolumn{12}{l}{Sm, Im}\\
\object{NGC 1156}&SmIV		&IBm(s)		&   3.3 & 42   	&25	&11.41	&372	&C &    25& 115	& Hunter et al. 2002	\\
\object{NGC 2366}&SBmIV-V	&IBm(s)		&   8.1 & 66   	&25	&10.95	&87	&I &    30& 120	& Hunter et al. 2002	\\
\object{NGC 4449}&SmIV		&IBm		&   6.2 & 45   	&45	&9.94	&211	&Z &    45& 135	& Hunter et al. 2002	\\
\noalign{\smallskip}
\multicolumn{12}{l}{S}\\
\object{NGC 1487}&S pec tides	&pec		&   3.3 & 50  	&55	&11.92$^*$&733	&G &\ldots& 160	& Aguero \& Paolantonio 1997\\
\object{NGC 3691}&S		&SBb?		&   1.3 & 42   	&15	&12.95	&997	&Z &   191& 100	& Blackman 1980b        \\
\object{NGC 3955}&S pec      	&S0/a pec       &   2.9 & 74    &165  	&12.03  &1515   &G &   166&  76	& Chromery 1974		\\
\noalign{\smallskip}
\multicolumn{12}{l}{SB}\\
\object{NGC 4670}&SB pec        &SB$^0$(s): pec	&   1.4 & 41   	&90	&12.89	&1112	&Z &    90&   0	& Sakka et al. 1973	\\
		 &		&		&       &       &	&    	&	&  &&		&			\\
\hline
\label{tab:catalog}
\end{longtable}
\begin{minipage}{23.5cm}
NOTES -- Data are taken from RC3 except for those marked with *, which are taken from LEDA.
Col.(2): morphological classification from RSA.
Col.(3): morphological classification from RC3.
Col.(4): major-axis isophotal diameter measured at the surface brightness 
         level $\mu_B$ = 25.0 mag arcsec$^{-2}$ from RC3.
Col.(5): inclination derived as $\cos^{2}{i}\,=\,(q^2-q_0^2)/(1-q_0^2)$.
         The observed axial ratio $q$ is taken from RC3 and the
         intrinsic flattening has been assumed following
         Guthrie (1992) with RC3 morphological classification.
Col.(6): major-axis position angle from RC3.
Col.(7): apparent total blue magnitude corrected for
         inclination, extinction and redshift from RC3.
Col.(8): optical recession velocity from RC3.
Col.(9): classification of the minor-axis velocity profiles as described in 
         Sect.\ref{sec:catalog}.
Col.(10): observed position angle which is closest to the galaxy major axis. 
Col.(11): observed position angle which is closest to the galaxy minor axis. 
Col.(12): reference for gas kinematics measured along the observed position angles.
\end{minipage}
\end{landscape}
\end{tiny}


\begin{thebibliography}{}

\bibitem[Afanazyev et al. 1992]{afa92} Afanasyev, V. L., Burenkov,
  A. N., Zasov, A. V., \& Silchenko, O. K. 1992, AZh, 69, 19

\bibitem[Aguero et al. 1997]{agu97} Aguero, E. L., \& Paolantonio,
  S. 1997, AJ, 114, 102

\bibitem[Athanassoula(1992)]{1992MNRAS.259..345A} Athanassoula, E.\
  1992, MNRAS, 259, 345

\bibitem[Beauvais et al. 2001]{bea01}Beauvais, C., \& Bothun, G. 2001,
  ApJS, 136, 41

\bibitem[Berman (2001)]{ber01} Berman, S. 2001, A\&A, 371, 476

\bibitem[Bertola \& Corsini (1999)]{ber99a} Bertola, F., \& Corsini,
  E.  M.  1999, in Galaxy Interactions at Low and High Redshift, ed.
  J.  Barnes, \& D. B. Sanders (Dordrecht: Kluwer Academic Press), IAU
  Symp. 186, 149

\bibitem[Bertola et al. 1989]{ber89} Bertola, F., Rubin, V. C., \&
  Zeilinger, W. W.  1989, ApJ, 345, L29

\bibitem[Bertola et al. 1991]{ber91} Bertola, F., Vietri, M., \& Zeilinger, W. W.
1991, ApJ, 374, L13


\bibitem[Bertola, Buson, \& Zeilinger(1992)]{1992ApJ...401L..79B}
  Bertola, F., Buson, L.~M., \& Zeilinger, W.~W.\ 1992a, ApJ, 401,
  L79

\bibitem[Bertola et al. 1992]{ber92} Bertola, F., Galletta, G., \&
  Zeilinger, W. W. 1992b, A\&A, 254, 89

\bibitem[Bertola et al.(1995)]{1995ApJ...448L..13B} Bertola, F.,
  Cinzano, P., Corsini, E.~M., Rix, H., \& Zeilinger, W.~W.\ 1995,
  ApJ, 448, L13

\bibitem[Bertola et al.(1998)]{1998ApJ...509L..93B} Bertola, F.,
  Cappellari, M., Funes, J.~G., Corsini, E.~M., Pizzella, A., \& Vega
  Beltr\'an, J.~C.\ 1998, ApJ, 509, L93

\bibitem[Bettoni et al. 1997]{bet97} Bettoni, D., \& Galletta,
  G. 1997, A\&AS, 124, 61

\bibitem[Bettoni et al. 1990]{bet90} Bettoni, D., Fasano, G., \&
  Galletta, G. 1990, AJ, 99, 1789

\bibitem[Blackman 1980]{bla80} Blackman, C. P. 1980a, MNRAS, 190, 459

\bibitem[Blackman 1980b]{bla80b} Blackman, C. P. 1980b, MNRAS, 191, 123

\bibitem[Blackmann et al. 1983]{bla83} Blackman, C. P., Wilson, A. S.,
  \& Ward, M. J. 1983, MNRS, 202, 1001

\bibitem[Burbidge et al. 1960]{bur60} Burbidge, E. M., Burbidge,
  G. R., \& Prendergast, K. H. 1960, ApJ, 132, 654

\bibitem[Burbidge et al. 1962]{bur62} Burbidge, E. M., Burbidge,
  G. R., \& Prendergast, K. H. 1962, ApJ, 136, 339

\bibitem[Burbidge et al. 1963]{bur63} Burbidge, E. M., Burbidge,
  G. R., \& Prendergast, K. H. 1963, ApJ, 137, 376

\bibitem[Buta 1987]{but87} Buta, R. 1987, ApJS, 64, 383

\bibitem[Caldwell 1984]{cal84} Caldwell, N. 1984, ApJ, 278, 96

\bibitem[Caon et al. 2000]{cao00} Caon, N., Macchetto, D., \&
  Pastorizia, M. 2000, ApJS, 127, 39


\bibitem[Carozzi-Meyssonnier 1979]{car79} Carozzi-Meyssonnier,
  N. 1979, A\&AS, 37, 529

\bibitem[Chincarini et al. 1967]{chi67} Chincarini, G., \& Walker,
  M. F. 1976, ApJ, 149, 487

\bibitem[Chromey et al. 1974]{chr74} Chromey, F. R. 1974, A\&A, 31,
  165

\bibitem[Cinzano et al.(1999)]{1999MNRAS.307..433C} Cinzano, P., Rix,
  H.-W., Sarzi, M., Corsini, E.~M., Zeilinger, W.~W., \& Bertola, F.\
  1999, MNRAS, 307, 433

\bibitem[Coccato et al. 2004]{coc04}  Coccato, L., Corsini, E. M., 
  Pizzella, A., \& Bertola, F. 2004, in preparation

\bibitem[Comte et al. 1979]{com79} Comte, G., Monnet, G., \& Rosado,
  M. 1979, A\&A, 72, 73

\bibitem[Corsini \& Bertola 1998]{cor98} Corsini, E. M., \& Bertola,
  F. 1998, JKPS, 33, 574

\bibitem[Corsini et al. 2002]{cor02} Corsini, E. M., Pizzella, A., \&
  Bertola, F. 2002, A\&A, 382, 488

\bibitem[Corsini et al. 2003]{cor03} Corsini, E. M., Pizzella, A.,
 Coccato, L., \& Bertola, F. 2003, A\&A, 408, 873

\bibitem[Corsini et al. (1999)]{cor1999} Corsini, E.\ M., Pizzella,
  A., Sarzi, M., Cinzano, P., Vega Beltr\'an, J. C., Funes, J.\ G.,
  Bertola, F., Persic, M., \& Salucci, P.  1999, A\&A, 342, 671

\bibitem[Corsini et al. (2000)]{cor00} Corsini, E.\ M., Pizzella,
  A., Sarzi, M., Vega Beltr\'an, J. C., Cappellari, M., Funes, J.\ G.,
  \& Bertola, F., 2000, in Dynamics of Galaxies: from the Early Universe to the 
  Present, ed. F. Combes, G. A. Mamon \& V. Charmandaris, ASP Conf. Ser. 197, 
  (ASP: San Francisco), 251

\bibitem[Danziger et al. 1981]{dan81} Danziger, I. J., Goss, W. M., \&
  Wellington, K. J. 1981, MNRAS, 195, 33

\bibitem[de Blok, McGaugh, \& Rubin(2001)]{2001AJ....122.2396D} de
  Blok, W.~J.~G., McGaugh, S.~S., \& Rubin, V.~C.\ 2001, AJ, 122,
  2396

\bibitem[Dehnen 1993]{www} Dehnen, W. 1993, MNRAS, 265, 250

\bibitem[Demoulin 1969]{dem69} Demoulin, M. 1969, ApJ, 157, 69

\bibitem[de Vaucouleurs 1991]{dev91} de Vaucouleurs, D.,de Vaucouleurs, A., 
 Corwin, H. G., Jr., Buta, R. J., Paturel, G. \& Fouque, P. 1991, 
Third Reference Catalog of Bright Galaxies (New York: Springer-Verlag) 

\bibitem[]{dez}de Zeeuw, P. T. \& Carollo, C. M. 1996, MNRAS, 281, 1333

\bibitem[de Zeeuw \& Franx(1989)]{1989ApJ...343..617D} de Zeeuw, T., \&
  Franx, M.\ 1989, ApJ, 343, 617

\bibitem[Duval 1981]{duv81} Duval, M. F., 1981, A\&A, 98, 352

\bibitem[Emsellem \& Arsenault(1997)]{1997A&A...318L..39E} Emsellem,
  E., \& Arsenault, R.\ 1997, A\&A, 318, L39

\bibitem[Fillmore, Boroson, \& Dressler(1986)]{1986ApJ...302..208F}
  Fillmore, J.~A., Boroson, T.~A., \& Dressler, A.\ 1986, ApJ, 302,
  208

\bibitem[Fisher 1997]{fis97} Fisher, D. 1997, AJ, 113, 950

\bibitem[Fisher et al. 1994]{fis94} Fisher, D., Illingworth, G., \&
  Franx, M. 1994, AJ, 107, 160

\bibitem[Frei et al. 1996]{fre96}Frei, Z., Guhathakurta, P., Gunn, J. E., \&
 Tyson, J. A. 1996, AJ, 111, 174

\bibitem[Fridman et al. 2001]{fri01} Fridman, A. M., Khoruzhii, O. V.,
  Lyakhovich, V. V., Sil'chenko, O. K., Zasov, A. V., Afanasiev,
  V. L., Dodonov, S. N., \& Boulesteix, J. 2001, A\&A, 371, 538

\bibitem[Friedli \& Benz(1993)]{1993A&A...268...65F} Friedli, D.~\&
  Benz, W.\ 1993, A\&A, 268, 65

\bibitem[Funes et al. 2002]{fun02} Funes, J. G., S. J, Corsini, E. M.,
  Cappellari, M., Pizzella, A., Vega Beltr\'an, J. C., Scarlata, C.,
  \& Bertola, F. 2002, A\&A, 388, 50

\bibitem[Gerhard \& Vietri(1986)]{1986MNRAS.223..377G} Gerhard, O.~E.,
  \& Vietri, M.\ 1986, MNRAS, 223, 377

\bibitem[Goad 1974]{goa74} Goad, J. W. 1974, ApJ, 192, 311

\bibitem[Haynes et al. 2000]{hay00} Haynes, M. P., Jore, K. P.,
  Barrett, E. A., Broeils, A. H., \& Murray, B. M. 2000, AJ, 120, 703

\bibitem[Higdon et al. 1980]{hig80} Higdon, J. L., Buta, R. J., \&
  Purcell, G. B. 1980, AJ, 115, 80

\bibitem[Hunter et al. 2002]{hun02} Hunter, D. A., Rubin, V. C.,
  Swaters, R. A., Sparke, L. S., \& Levine, S. E. 2002, ApJ, 580, 194

\bibitem[Jarrett et al. 2000]{jar00}Jarrett, T. H., Chester, T., Cutri, R.,
 Schneider, S., Skrutskie, M., \& Huchra, J. P. 2000, AJ, 119, 2498

\bibitem[Jimenez et al. 2003]{jim03} Jimenez, R., Verde, L., \& Oh, S. P. 
2003, MNRAS, 339, 243

\bibitem[Jogee et al. 1999]{jog99} Jogee, S., Kenney, J. D. P., \&
  Smith, B. J. 1999, ApJ, 526, 665

\bibitem[Jore et al. 1996]{jor96} Jore, K. P., Broelis, A. H., \&
  Haynes, M. P. 1996, AJ, 112, 438

\bibitem[Kenney et al. 1996]{ken96} Kenney, J. D. P., Koopmann, R. A.,
  Rubin, V. C., \& Young, J. S. 1996, AJ, 111, 152

\bibitem[Kent(1986)]{1986AJ.....91.1301K} Kent, S.~M.\ 1986, AJ, 91,
  1301

\bibitem[Keppel at al. 1991]{kep91} Keppel, J. W., Dettmar, R. J.,
 Gallagher, J. S., III, \& Roberts, M. S. 1991, ApJ, 374, 507

\bibitem[Kuijken, Fisher, \& Merrifield(1996)]{1996MNRAS.283..543K}
  Kuijken, K., Fisher, D., \& Merrifield, M.~R.\ 1996, MNRAS, 283,
  543

\bibitem[Kuno et al. 1997]{kin97} Kuno, N., Tosaki, T., Nakai, N., \&
 Nishiyama, K. 1997, PASJ, 49, 275

\bibitem[Kyazumov 1980]{kya80} Kyazumov, G. A. 1980, SvA, 6, L398

\bibitem[Lake et al. 1986]{lak86} Lake, G, \& Dressler, A. 1986, ApJ,
  310, 605

\bibitem[Lindblad et al. 1996]{lin96} Lindblad, P. O., Hjelm, M.,
  Hoegbom, J., Joersaeter, S., Lindblad, P. A. B., \& Santos-Lleo,
  M. 1996, A\&AS, 120, 403

\bibitem[Marquez et al. 1996]{mar96} M\'arquez, I., \& Moles, M. 1996,
  A\&A, 120, 1

\bibitem[Marquez et al. 1998]{mar98} M\'arquez, I., Boisson, C., Durret,
  F., \& Petitjean, P. 1998, A\&A, 333, 459

\bibitem[Marquez et al. 2002]{mar02} M\'arquez, I., Masegosa, J., Moles,
  M., Varela, J., Bettoni, D., \& Galletta, G. 2002, A\&A, 
  393, 389

\bibitem[McGaugh, Rubin, \& de Blok(2001)]{2001AJ....122.2381M}
  McGaugh, S.~S., Rubin, V.~C., \& de Blok, W.~J.~G.\ 2001, AJ, 122,
  2381

\bibitem[Moellenhoff 1982]{moe82} Moellenhoff, C. 1982, A\&A, 108, 130

\bibitem[Moore 1994]{moo94} Moore, B. 1994, Nature, 370, 629

\bibitem[Mulder et al. 1996]{mul96} Mulder, P. S., \& van Driel,
 W. 1996, A\&A, 309, 403

\bibitem[Munoz Tunon 1987]{mub87}Munoz Tunon, C., Beckman, J., \&
  Prieto, M. 1987, Rev. Mex. Astr. Astrofis., 14, 144

\bibitem[Osterbrock et al.(1996)]{1996PASP..108..277O} Osterbrock,
  D.~E., Fulbright, J.~P., Martel, A.~R., Keane, M.~J., Trager, S.~C.,
  \& Basri, G.\ 1996, PASP, 108, 277

\bibitem[Pellet 1976]{pel76} Pellet, A. 1976, A\&A, 50, 421

\bibitem[Perez 2000]{per00} P\'erez, E., M\'arquez, I., Marrero, I.,
  Durret, F., Gonz\'alez Delgrado, R. M., Masegosa, J., Maza, J., \&
  Moles, M. 2000, A\&A. 353, 893

\bibitem[Persic, Salucci, \& Stel(1996)]{1996MNRAS.281...27P} Persic,
  M., Salucci, P., \& Stel, F.\ 1996, MNRAS, 281, 27

\bibitem[Peterson 1980]{pet80b} Peterson, C. J. 1980, PASP 92, 397

\bibitem[Peterson et al. 1980]{pet80} Peterson, C. J., \& Huntley,
  J. M. 1980, ApJ, 242, 913

\bibitem[Peterson et al. 1978]{pet78} Peterson, C. J., Roberts, M. S.,
  Rubin, V. C., \& Ford, W. K., Jr. 1978, ApJ, 226, 770

\bibitem[Pizzella et al. 2002]{piz02} Pizzella, A., Corsini, E. M.,
  Morelli, L., Sarzi, M., Scarlata, C., Stiavelli, M.,\& Bertola,
  F. 2002, ApJ, 573, 131

\bibitem[Prada et al. 1998]{pra98} Prada, F., Gutierrez, C. M., \&
  McKeith, C. D. 1998, ApJ, 495, 765

\bibitem[Reshetnikov et al. 1994]{res94} Reshetnikov, V. P., \&
  Combes, F. 1994, A\&A, 291, 57

\bibitem[Rubin et al. 1964]{rub64} Rubin, V. C., Burbidge, E. M., Burbidge, G. R., \&
 Prendergast, K. H. 1964, ApJ, 140, 80

\bibitem[Rubin 1980]{rub80} Rubin, V. C. 1980, ApJ, 238, 808

\bibitem[Rubin 1994]{rub94} Rubin, V. C. 1994a, AJ, 107, 173

\bibitem[Rubin 1994b]{rub94b} Rubin, V. C. 1994b, AJ, 108, 456

\bibitem[Rubin et al. 1970]{rub70} Rubin, V. C., \& Ford, W. K.,
  Jr. 1970, ApJ, 159, 379

\bibitem[Rubin et al. 1965]{rub65} Rubin, V. C., Burbidge, E. M.,
  Burbidge, G. R., \& Prendergast, K. H. 1965, ApJ, 141, 885

\bibitem[Rubin et al. 1977]{rub77} Rubin, V. C., Thonnard, N., \&
  Ford, W. K., Jr. 1977, ApJ, 217, L1

\bibitem[Rubin et al. 1979]{rub79} Rubin, V. C., Roberts, M. S., \&
  Ford, W. K., Jr. 1979, ApJ, 230, 35

\bibitem[Rubin et al. 1980]{rub80b} Rubin, V. C., Thonnard, N., \&
  Ford, W. K., Jr. 1980, ApJ, 238, 471

\bibitem[Rubin et al. 1999]{rub99} Rubin, V. C., Waterman, A. H., \&
  Kenney, J. D. P. 1999, AJ, 118, 236

\bibitem[Ryder et al. 1998]{ryd98}Ryder, S. D., Zasov, A. V.,
  Sil'Chenko, O. K., McIntyre, V. J., \& Walsh, W. 1998, MNRAS, 293,
  411

\bibitem[Sandage \& Tammann 1981]{san81} Sandage, A. \& Tammann,
  G. A. 1981, A Revised Shapley-Ames Catalog of Bright Galaxies
  (Washington D.C.: Carnegie Institution of Washington)

\bibitem[Sandqvist et al. 1989]{san89} Sandqvist, A., Elfhag, T., \&
  Lindblad, P. O. 1989, A\&A, 218, 39

\bibitem[Sakka et al. 1973]{sak73} Sakka, K., Oka, S., \& Wakamatsu,
  K. 1973, PASJ, 25, 317

\bibitem[Sarzi 2003]{sar03} Sarzi, M. 2003, in Coevolution of Black
  Holes and Galaxies, ed. L. C. Ho (Cambridge:Cambdrige Univ. Press),
  in press 

\bibitem[Sarzi et al. (2000)]{sar2000} Sarzi, M., Corsini, E.\ M.,
  Pizzella, A., Vega Beltr{\' a}n, J.\ C., Cappellari, M., Funes, J.\ 
  G., \& Bertola, F.\ 2000, A\&A, 360, 439

\bibitem[Schechter et al. 1978]{sch78} Schechter, P.~L., \& Gunn,
  J. E. 1978, AJ, 83, 1360

\bibitem[Schulz et al. 1994]{sch94} Schulz, H., Knake, A., \&
  Schmidt-Kaler, T. 1994, A\&A, 288, 425

\bibitem[Sharples et al. 1983]{sha83} Sharples, R. M., Carter, D.,
  Hawarden, T.G., \& Longmore, A. J.  1983, MNRAS, 202, 37

\bibitem[Silchenko et al 1997]{sil97}Sil'chenko, O. K., Vlasyuk, V. V., Burenkov, A. N. 1997, A\&A, 326, 941

\bibitem[Sofue \& Rubin(2001)]{2001ARA&A..39..137S} Sofue, Y.~\&
  Rubin, V.\ 2001, ARA\&A, 39, 137

\bibitem[Sofue et al. 1999]{sof99} Sofue, Y., Tutui, Y., Honma, M.,
  Tomita, A., Takamiya, T., Koda, J., \& Takeda, Y. 1999, ApJ, 523, 136

\bibitem[Steiman-Cameron, Kormendy, \&
  Durisen(1992)]{1992AJ....104.1339S} Steiman-Cameron, T.~Y.,
  Kormendy, J., \& Durisen, R.~H.\ 1992, AJ, 104, 1339

\bibitem[Storchi-Bergmann 1996]{qwqw} Storchi-Bergmann, T.,
  Redrigues-Ardila, A., Schmitt, H.~R., Wilson, A.~S., Baldwin,
  J.~A. 1996, ApJ, 472, 83


\bibitem[va der Kruit 1976]{van76} van der Kruit, P. C. 1976, A\&A,
  49,161

\bibitem[van Gorkom et al. 1990]{van90} van Gorkom, J. H., van der
 Hulst, J. M., Haschick, A. D., \& Tubbs, A. D. 1990, AJ, 99, 1781

\bibitem[Vega Beltran et al. 1998]{veg98} Vega Beltran, J. C.,
  Zeilinger, W. W., Amico, P., Schultheis, M., Corsini, E. M., Funes,
  J. G., Beckman, J., \& Bertola, F. 1998, A\&AS, 131, 105

\bibitem[Veilleux et al. 1999]{vei99} Veilleux S., Bland-Hawthorn, J.,
  \& Cecil, G. 1999, AJ, 118, 2108

\bibitem[Walker 1989]{wal89} Walker, M. F. 1989, PASP, 101, 333

\bibitem[Wilke et al. 1999]{wil99} Wilke, K., M\"ollenhoff, C., \&
  Matthias, M. 1999, A\&A, 344, 787

\bibitem[Wilke et al. 2000]{wil00} Wilke, K., M\"ollenhoff, C., \&
  Matthias, M. 2000, A\&A, 361, 507

\bibitem[Winge et al. 1999]{win99} Winge, C., Axon, D. J., Macchetto,
  F. D., Capetti, A., \& Marconi, A. 1999, ApJ, 519, 134

\bibitem[Whitmore et al. 1990]{whi90} Whitmore, B. C. et al.  1990,
  AJ, 100, 1489

\bibitem[Zeilinger et al. 1990]{zei90} Zeilinger, W. W, Galletta, G.,
  \& Madsen, C. 1990, MNRAS, 246, 324

\bibitem[Zeilinger et al. 2001]{zei01} Zeilinger, W. W., Vega
  Beltr\'an, J. C., Rozas, M., Beckman, J. E., Pizzella, A., Corsini,
  E. M., Bertola, F.  2001, Ap\&SS, 276, 643

\end{thebibliography}
\end{document}